\documentclass[sigconf]{acmart}

\usepackage{nicefrac}
\usepackage{siunitx}
\usepackage{array,framed}
\usepackage{booktabs}
\usepackage{
  color,
  epsfig,
  wrapfig
}

\usepackage{textcomp,amssymb}
\usepackage{setspace}
\usepackage{latexsym,fancyhdr,url}
\usepackage{enumerate}
\usepackage{algorithm2e}
\usepackage{algpseudocode}
\usepackage{xparse} 
\usepackage{xspace}
\usepackage{multirow}
\usepackage{csvsimple}
\usepackage{balance}
\usepackage{ulem} 
\usepackage{graphicx,subcaption}
\usepackage{stfloats}
\usepackage{caption}

\usepackage{
  tikz,
  pgfplots,
  pgfplotstable
}
\usepackage{hyperref}

\usepackage{lineno}
\usepackage{amsmath}
\usepackage{nomencl}
\usepackage{ulem}
\usepackage{ifthen}
\usepackage{makecell}

\usetikzlibrary{
  shapes.geometric,
  arrows,
  external,
  pgfplots.groupplots,
  matrix
}
\usepackage{xcolor}

\AtBeginDocument{%
  }

\newcommand{\zyz}[1]{ \textcolor{black}{#1}}
\newcommand{\projname}{UWBAD}
\begin{document}

\title{UWBAD: Towards Effective and Imperceptible Jamming Attacks\\ Against UWB Ranging Systems with COTS Chips}

\author{Yuqiao Yang}
\affiliation{%
  \institution{\textit{UESTC}}
  \city{Shenzhen}
  \country{China}
}
\email{yangyuqiao@gogobyte.com}

\author{Zhongjie Wu}
\affiliation{%
  \institution{\textit{GoGoByte Technology}}
  \city{Beijing}
  \country{China}
}
\email{wuzhongjie@gogobyte.com}

\author{Yongzhao Zhang}
\authornote{Corresponding authors.}
\affiliation{%
  \institution{\textit{UESTC}}
  \city{Chengdu}
  \country{China}
}
\email{zhangyongzhao@uestc.edu.cn}

\author{Ting Chen}
\authornotemark[1]
\affiliation{%
  \institution{\textit{UESTC}}
  \city{Chengdu}
  \country{China}
}
\email{chenting19870201@163.com}

\author{Jun Li}
\affiliation{%
  \institution{\textit{GoGoByte Technology}}
  \city{Beijing}
  \country{China}
}
\email{lijun@gogobyte.com}

\author{Jie Yang}
\affiliation{%
  \institution{\textit{UESTC}}
  \city{Chengdu}
  \country{China}
}
\email{jie.yang@uestc.edu.cn}

\author{Wenhao Liu}
\affiliation{%
  \institution{\textit{GoGoByte Technology}}
  \city{Beijing}
  \country{China}
}
\email{lwh.scu@gmail.com}

\author{Xiaosong Zhang}
\affiliation{%
  \institution{\textit{UESTC}}
  \city{Chengdu}
  \country{China}
}
\email{johnsonzxs@uestc.edu.cn}

\author{Ruicong Shi}
\affiliation{%
  \institution{\textit{GoGoByte Technology}}
  \city{Beijing}
  \country{China}
}
\email{shiruicong@gogobyte.com}

\author{Jingwei Li}
\affiliation{%
  \institution{\textit{UESTC}}
  \city{Chengdu}
  \country{China}
}
\email{jwli@uestc.edu.cn}

\author{Yu Jiang}
\affiliation{%
  \institution{Tsinghua University}
  \city{Beijing}
  \country{China}
}
\email{jiangyu198964@126.com}

\author{Zhuo Su}
\affiliation{%
  \institution{Tsinghua University}
  \city{Beijing}
  \country{China}
}
\email{suzcpp@gmail.com}

\acmDOI{10.1145/3658644.3670349}

\copyrightyear{2024}
\acmYear{2024}
\setcopyright{rightsretained}

\acmConference[CCS '24] {Proceedings of the 2024 ACM SIGSAC Conference on Computer and Communications Security}{October 14--18, 2024}{Salt Lake City, UT, USA.}
\acmISBN{979-8-4007-0636-3/24/10}
\renewcommand{\shortauthors}{Yang et al.}
\acmBooktitle{Proceedings of the 2024 ACM SIGSAC Conference on Computer and Communications Security (CCS '24), October 14--18, 2024, Salt Lake City, UT, USA}


\begin{abstract}
UWB ranging systems have been adopted in many critical and security sensitive applications due to its precise positioning and secure ranging capabilities. 
We present a practical jamming attack, namely \projname{}, against commercial UWB ranging systems, which exploits the vulnerability of the adoption of the normalized cross-correlation process in UWB ranging and can \zyz{selectively} and quickly block ranging sessions without prior knowledge of the configurations of the victim devices, potentially leading to severe consequences such as property loss, \zyz{unauthorized access}, or vehicle theft. \projname{} achieves more effective and less imperceptible jamming due to: (i) it efficiently blocks every ranging session by leveraging the field-level jamming, thereby exerting a tangible impact on commercial UWB ranging systems, and (ii) \zyz{the compact, reactive, and selective system design based on COTS UWB chips, making it affordable and less imperceptible.} We successfully conducted real attacks against commercial UWB ranging systems from the three largest UWB chip vendors on the market, e.g., Apple, NXP, and Qorvo. We reported our findings to Apple, related Original Equipment Manufacturers (OEM), and the Automotive Security Research Group, triggering internal security incident response procedures at Volkswagen, Audi, Bosch, and NXP. 
As of the writing of this paper, \zyz{the related OEM has acknowledged this vulnerability in their automotive systems and has offered a $\$5,000$ reward as a bounty.}
\end{abstract}




\maketitle


\section{Introduction}
\label{sec:intro}


Ultra Wide-Band (UWB) ranging systems adhering to IEEE 802.15.4z standard use a large bandwidth exceeding 500MHz and unforgeable ranging packet design for precise and secure ranging. It has been widely used in many industry applications, such as indoor localization~\cite{grosswindhager2019snaploc,zafari2019survey}, manufacturing and logistics~\cite{elsanhoury2022precision,zhao2020joint}, asset tracking~\cite{Apple2,shyam2022uwb}, and access and authorization systems~\cite{PKE16,PKE23,PKE59,pke111}.

Precise ranging in UWB systems is achieved through the measurement of the Time-of-Flight (ToF) of radio waves from one device to another and back. The ToF is then multiplied by the speed of light to determine the distance between the two devices. The broad spectrum range of UWB in the frequency domain generates extremely short pulses in the time domain, typically on the order of several nanoseconds~\cite{leu2022ghost}. This narrow pulse width renders UWB practical for accurate ToF measurement, even in the presence of severe multipath effects~\cite{firaAccurateRanging}. 
By continuously starting the ranging sessions (transmitting several ranging packets for timestamp measurement for each session), 
the UWB ranging system can keep tracking and updating the precise position of the target device.

\begin{figure*}[htbp]
    \centering
    \begin{subfigure}[b]{0.33\linewidth} 
        \centering
        \includegraphics[width=0.92\textwidth]{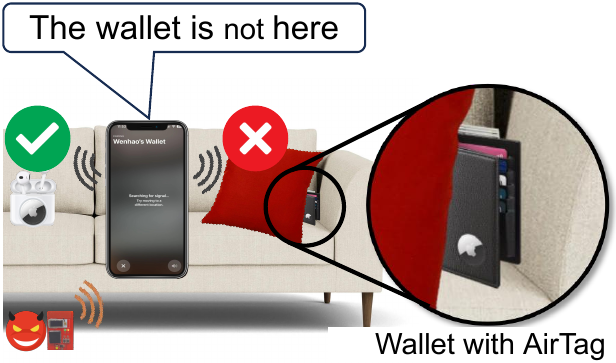}
        \caption{\zyz{iPhone cannot find the wallet.}}
        \label{fig:intro_tag}
    \end{subfigure}
    \vspace{5pt}
    \hfill 
    \begin{subfigure}[b]{0.33\linewidth}
        \centering
        \includegraphics[width=0.78\textwidth]{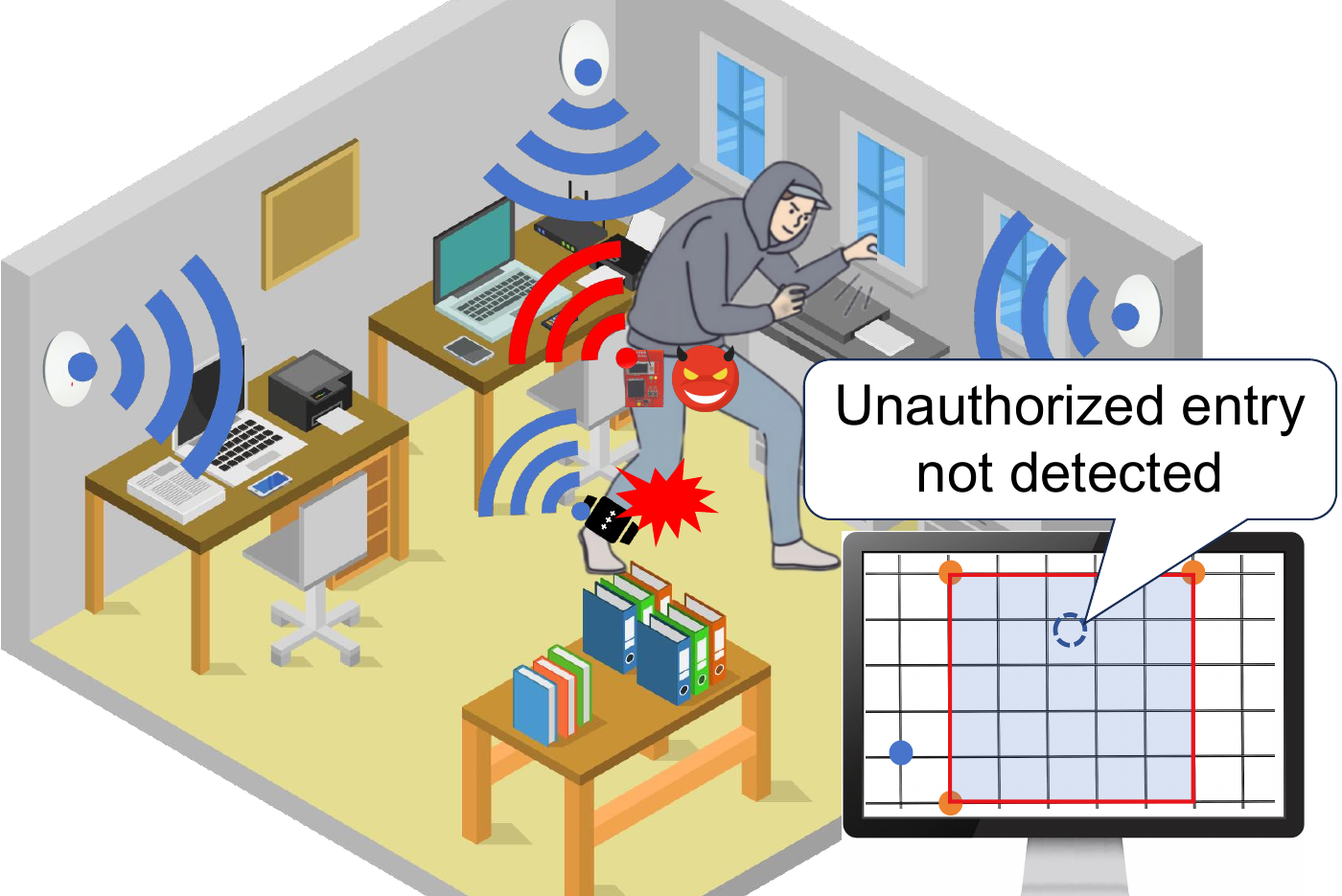}
        \caption{\zyz{Unauthorized entry does not trigger alerts.}}
        \label{fig:intro_localization}
    \end{subfigure}
    \vspace{5pt}
    \hfill
    \begin{subfigure}[b]{0.33\linewidth}
        \centering
        \includegraphics[width=0.92\textwidth]{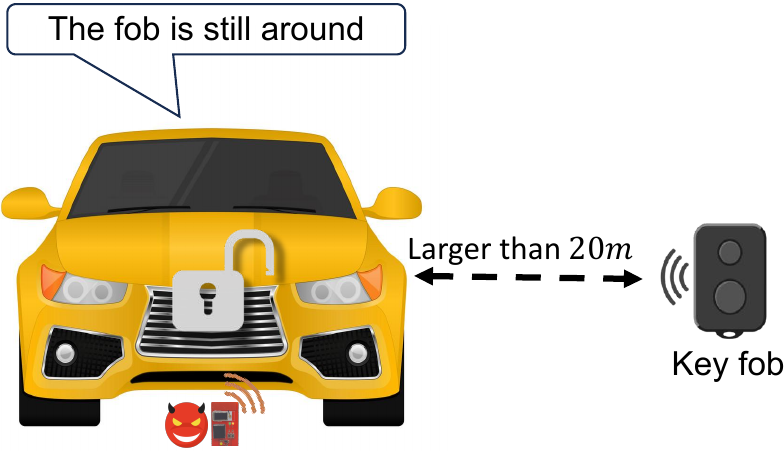}
        \caption{\zyz{Unauthorized vehicle access and theft.}}
        \label{fig:intro_car2}
    \end{subfigure}
    \vspace{-25pt}
    \caption{\zyz{Attack scenarios of \projname{}. (a) \textbf{Asset Tracking:} An adversary can 
    fail the UWB ranging between an iPhone and the wallet (attached with an AirTag) and easily steal the wallet, while the other iPhone-AirTag pair remains unaffected.
    (b) \textbf{Indoor Localization:} An adversary wearing a non-removable localization tag can freely enter or leave the restricted areas with \projname{}.
    (c) \textbf{Vehicle Theft:} The adversary disrupts the ranging process of a passive keyless entry and start (PKES) system of vehicles, such that the car believes that the key fob is still around, even when the user has walked a long distance.}
    }\label{fig:intro}
    \vspace{-10pt}
\end{figure*}

Secure ranging is crucial for the adoption of UWB systems in security-sensitive applications. Several attacks against UWB ranging systems have been studied in the literature, such as jamming attacks~\cite{dospaper}, relay attacks~\cite{singh2017uwb}, and distance reduction attacks~\cite{leu2022ghost}. For jamming attacks, UWB is known to be immune to narrow-band jamming~\cite{immunejamming}, whereas the full-band jamming necessitates expensive, bulky, and customized hardware~\cite{keysight}.\zyz{More importantly, full-band jamming lacks the ability to selectively attack  a specific device without impacting other devices, rendering it more perceptible and detectable in real-world scenarios.}
For replay attacks, the recently implemented encrypted scrambled timestamp sequence (STS) field in 4z standard can effectively reject replayed packets, thus safeguarding against these types of attacks. 
The most recent distance reduction attack, Ghost Peak~\cite{leu2022ghost}, can be applied to UWB with 4z standard. It, however, is less efficient as the attack success rate is only $4\%$. Moreover, the reduced distance measurement can be easily removed by filtering~\cite{welch1995introduction, justusson2006median} or be rejected by using sub-template verification~\cite{joo2023protecting}, rending this attack less effective.
Therefore, effective yet efficient jamming attacks against commercial UWB ranging systems adhering to the 4z standard remain underexplored.

In this paper, we propose \projname{} (\underline{\textit{UWB}} \underline{\textit{A}}ccurate \underline{\textit{D}}eafening), a reactive jamming attack developed based on Commercial-Off-The-Shelf (COTS) UWB chips to fail UWB ranging systems. Our approach is more effective and less perceptible in completely blocking the UWB ranging systems even with the 4z standard. First, \projname{} is highly effective as it can efficiently jam the ranging system by blocking every ranging session of the victim UWB ranging system. Unlike distance reduction attacks, our jamming attack does not manipulate the measured distance but rather prevents every single distance updating by blocking ranging sessions.
\zyz{Second, \projname{} launches reactive and selective jamming by leveraging low-cost and compact COTS UWB chips.}
It listens to the communication and launch attacks reactively relying only on the limited control of the well-encapsulated chips, without using the full-band noise flooding that is easily detectable. \zyz{Moreover, \projname{} facilitates targeted attacks on victim devices without causing disruption to neighboring devices, making our attacks less perceptible.}

The basic idea of our proposed \projname{} is to exploit the vulnerability of the packet detection scheme designed in the UWB ranging system. In particular, as it is difficult to directly detect the UWB pulses under low SNR and low duty cycle (less than $1\%$~\cite{qian2022overview}), the UWB ranging system adopts the normalized cross-correlation (NCC) to estimate the channel impulse response (CIR) by correlating the received signal with a local template. We analyzed this process and found that the NCC only compares the similarity of two signals regardless of the signal power. This implies that we could distort the received packet in each ranging session to reduce the similarity below the threshold, thus jamming the ranging system. 
Still, realizing such an attack needs to overcome the following two challenges. 

\textit{First, how to maximize the efficiency of jamming each ranging session?}
As the UWB ranging systems keep tracking the position of the target device by continuously starting the ranging sessions, if only a fraction of the sessions are blocked, the ranging system can still function as it merely diminishes the efficiency of the updating rate of the ranging system. It is thus critical to maximize the effectiveness of jamming each session to block all the ranging updates. Although the jamming packets generated by COTS UWB chips naturally occupy the entire UWB communication bandwidth, achieving pulse-level jamming or manipulation is highly challenging and inefficient in practice. However, the above-mentioned normalized cross-correlation (NCC) at the receiver for CIR detection opens a door for us to perform field-level jamming and efficiently disrupt packet detection. 
To be specific, we found that attacking the SYNC field (i.e., distorting the SYNC field by jamming) instead of the STS field can maximize efficiency as the SYNC field adopts a much shorter code with tens of repetition, reducing the correlation property of the SYNC field. Therefore, injecting energy into the SYNC field can be more effective and efficient in reducing the maximum correlation below the threshold, blocking each ranging update, and jamming the victim ranging systems.

\textit{Second, how to inject the jamming packets with COTS UWB chips without knowing the physical-layer structure of the ranging packet?} 
Consistently and accurately superimposing the attack signal onto the SYNC field of the ranging packet requires knowledge of the precise timing to initiate the jamming attack.
However, retrieving the timestamp of the received ranging packet and the SYNC field demands awareness of the physical layer structure of the target packet, which is product-specific and not publicly available. 
While it is theoretically possible to pre-test or exhaustively configure all UWB-enabled products on the market, this is impractical and inefficient, especially for newly released, unseen products.
Therefore, \projname{} should be able to (i) quickly sniff the physical-layer structure of the ranging packet from the victim devices and (ii) predict the correct time to achieve the superposition of SYNC fields and initiate the jamming attack. To address these challenges, we leverage the responses of COTS UWB chips as hints, dividing the sniffing process into stages, and significantly reducing the potential search space. This reduces the preparation time for packet sniffing from a maximum of $146$ hours (brute-force approach) to an average of around $22.46$ seconds. Subsequently, we can record the timestamp of each packet and compute the time interval between two packets which is, surprisingly, fixed. 
Thus, we can effectively predict the next packet arrival time to launch the jamming attack at the right time with high efficiency. 

Disrupting UWB's ranging sessions has significant implications for various commercial products in the market, and Figure~\ref{fig:intro} depicts \zyz{three} examples we have conducted in our case study evaluation.
\zyz{(i) Apple's recently launched AirTag~\cite{Apple2} employs UWB technology to track the location of personal items such as wallets and earphones. As illustrated in Figure~\ref{fig:intro_tag}, a tiny and compact \projname{} device can be discreetly placed to selectively block the ranging packets from the AirTag attached to a wallet, while leaving the AirTag on the earphones unaffected. Consequently, the user can still locate the earphones using an iPhone but will be unable to find his/her wallet. This makes the \projname{} attack imperceptible to users. As a result, an adversary could easily steal the wallet without being noticed.}
\zyz{(ii) UWB-based indoor positioning systems are widely used for access control in settings such as warehouses~\cite{uwbwarehouse}, jails~\cite{uwbprison}, and hospitals~\cite{uwbhospital}. These systems are critical for preventing unauthorized access, asset theft, or prison breaks. As illustrated in Figure~\ref{fig:intro_localization}, an adversary equipped with a non-removable localization tag, when restricted from certain areas, can carry a \projname{} device to disrupt the UWB ranging sessions. This interference causes the localization system to fail in updating his/her real-time positions, allowing the adversary to freely enter or exit restricted areas undetected.}
(iii) Nowadays, many high-end vehicles feature the Passive Keyless Entry and Start System (PKES)~\cite{BWMUWB1, TeslaUWB}, enabling owners to unlock/lock and start the car without taking out the key fob. However, as illustrated in Figure~\ref{fig:intro_car2}, we use \projname{} device to thwart the car from being locked even after the owner leaves. Then, the adversary can start the car as the ranging system believes the key fob remains nearby. This situation arises because the UWB ranging system of PKES continues reading the old distance data in the absence of updates.
We reported our findings to Apple, related OEM, and the Automotive Security Research Group (ASRG), triggering internal security incident response procedures at Volkswagen, Audi, Bosch, and NXP. As of the writing of this paper,\zyz{related OEM has acknowledged this vulnerability as 'Critical' in their automotive systems and has offered a $\$5,000$ reward as a bounty.} Bosch and NXP are actively engaged in discussions with us to explore potential solutions to fix the vulnerability that we discovered in the UWB pulses detection scheme of the UWB ranging system.

Our hardware prototype, empowered by \projname{}, is built upon the COTS UWB chip DW3210~\cite{dw3210}. This device is compact and easy to reproduce, facilitating imperceptible and reliable deployment for attacks. \zyz{We conducted comprehensive experiments to evaluate the performance of \projname{} across various commercial devices. We first evaluated the attack effectiveness of \projname{} using a commercial UWB development kit (developed based on DW3210 chips), chosen for its customizable physical layer parameters. Then we conducted three case studies using commercial products, including the iPhone 14 and AirTags in asset tracking scenario, the indoor positioning system, and commercial vehicles and digital keys (PKES) system.} The above victim devices are from the three major vendors in the UWB chip market: Apple (iPhone 14 and AirTag), Qorvo (indoor positioning system), and NXP (key fob and car). \projname{} can completely fail the ranging sessions of all the aforementioned commercial devices, demonstrating robust performance in real-world tests.


Our contributions are summarized as follows: 
\begin{itemize}
    \item To the best of our knowledge, we are the first to exploit the vulnerability of adopting the NCC process in UWB ranging for jamming attacks.
    
    

    \item We leverage NCC in the SYNC field for packet detection to effectively \zyz{and selectively} block every ranging session of commercial UWB ranging systems.
    

    \item We implement \projname{} on a hardware prototype with COTS UWB chips, which can quickly prepare and launch attacks against unseen UWB products automatically. 

    \item We conduct comprehensive experiments to evaluate the effectiveness of attacking commercial UWB ranging systems, including the three largest vendors in the market of UWB chips, i.e., Apple, Qorvo, and NXP.
\end{itemize}

The rest of this work is organized as follows. We introduce the background in Sec.~\ref{sec:background} and the threat model in Sec.~\ref{sec:threat_model}. Then we discuss the core idea of jamming attack in Sec.~\ref{sec:core_idea}, and elaborate on how to implement \projname{} for the practical jamming attack in Sec.~\ref{sec:design}. Next, we evaluate the performance in Sec.~\ref{sec:eval}, before discussing the potential countermeasures, the feasibility of attacking the upcoming 4ab standard, and future works in Sec.~\ref{sec:Discuss}. Finally, we review the related work in Sec.~\ref{sec:related} and conclude in Sec.~\ref{sec:conclusion}.

\section{Background}
\label{sec:background}

The 4z standard delineates two modes of UWB ranging: Low-Rate Pulse Repetition Frequency (LRP) and High-Rate Pulse Repetition Frequency (HRP). LRP mode caters to low-power applications, while HRP mode facilitates faster data transmission, making HRP UWB more prevalent in commercial products. Consequently, our focus in this paper primarily revolves around HRP UWB. In this section, we will initially provide a brief overview of the fundamentals of HRP UWB ranging, followed by a discussion on how the latest 4z standard achieves secure ranging. 

\subsection{Fundamentals of HRP UWB Ranging}
UWB ranging systems achieve precise ranging by accurately measuring the timestamps of each UWB packet. Below, we delve into how UWB derives distance through packet transmission and discuss why UWB can ascertain accurate timestamps to achieve precise ranging.

\begin{figure}
    \centering
    \includegraphics[width=0.7\linewidth]{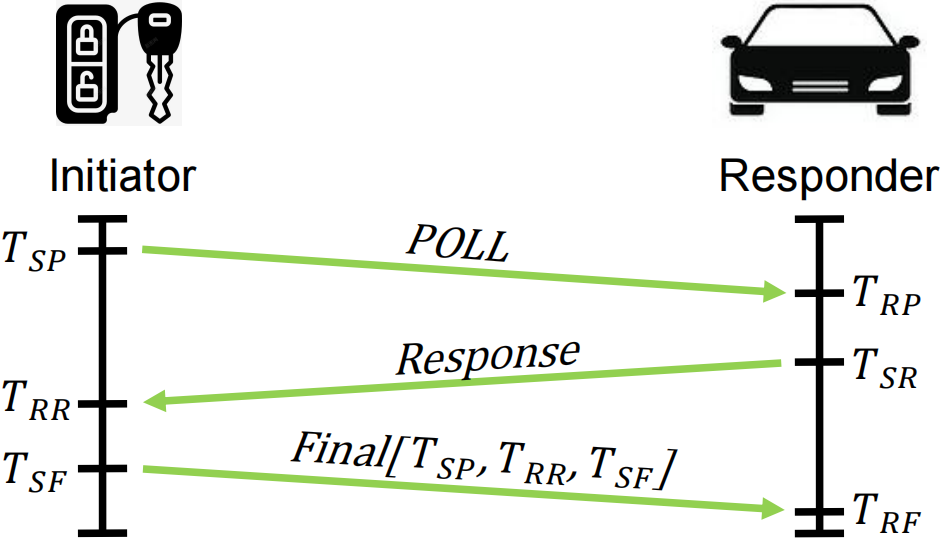}
    \vspace{-5pt}
    \caption{Double-sided two-way ranging (DS-TWR).}
    \label{fig:two_way_ranging}
    \vspace{-12pt}
\end{figure}

\textbf{Two-way Ranging (TWR):}
UWB employs two-way ranging (TWR), wherein messages are transmitted multiple times between the initiator and responder. As depicted in Figure~\ref{fig:two_way_ranging}, the initiator dispatches a poll message (the first packet) to the responder, recording the time $T_{SP}$ upon transmission. Upon receiving the poll message, the responder notes the time $T_{RP}$ of reception and subsequently dispatches a response message (the second packet) back to the initiator. Upon recording the arrival time $T_{RR}$, the initiator computes the distance using the following formula:
\begin{equation}
    d = c\Delta t = \frac{c}{2}[T_{RR}-T_{SP} - (T_{SR} - T_{T_{RP}})]
\end{equation}
where $\Delta t$ represents the ToF of the UWB signal between the initiator and the responder, and $c$ denotes the speed of light. This method, termed single-sided two-way ranging (SS-TWR), is susceptible to inaccuracies due to clock drift in both the initiator and responder. To mitigate this issue, the initiator dispatches an additional final message (the third packet) to the responder. Similarly, they record the time $T_{SF}$ of message transmission and the time $T_{RF}$ of message reception. Subsequently, the responder can rectify the clock drift error by computing the distance using the following formula:
\begin{equation}\label{eq:dstwr}
    d = c\Delta t = \frac{c}{4}[T_{RR}-T_{SP} - (T_{SR} - T_{T_{RP}}) + T_{RF} - T_{SR} - (T_{SF} - T_{RR})]
\end{equation}
This method, known as double-sided two-way ranging (DS-TWR), offers enhanced ranging accuracy compared to SS-TWR by mitigating the impact of clock drift. This encapsulates the core concept of DS-TWR, while for a deeper understanding of the DS-TWR ranging technique, we refer readers to \cite{neirynck2016alternative}, as the ranging principle is not the focus of this paper. The 4z standard encompasses both SS-TWR and DS-TWR. SS-TWR is more power-efficient than DS-TWR and is suitable for applications requiring low power consumption. However, DS-TWR is more prevalent in commercial products due to its superior ranging accuracy. \projname{} is applicable to both SS-TWR and DS-TWR.

\textbf{Normalized Cross-Correlation (NCC):}
The precise ToF measurement of UWB ranging systems relies on NCC. To circumvent interference with other radio links, the Federal Communications Commission (FCC) imposes strict limitations on the transmission power of UWB devices to $-41.3dBm/MHz$~\cite{krebesz2017use}, which is considerably lower than other wireless technologies such as WiFi and Bluetooth, and slightly above the noise floor. The low power density results in UWB pulses being modulated very sparsely in the time domain. Consequently, directly detecting UWB pulses with such low Signal-to-Noise Ratio (SNR) and low duty cycle is challenging. To address this, the receiver typically employs NCC to estimate the CIR, thereby consolidating the power of multiple pulses to generate a stronger peak. The basic cross-correlation function is defined as follows:
\begin{equation}
    R_{x,y}(\tau) = \sum_{l=0}^{N-1} x(t) \times y(t+\tau)
\end{equation}
where $x(t)$ represents the received signal, $y(t)$ denotes the local template for both the initiator and responder, $\tau$ signifies the time shift, and $N$ represents the signal length. As a thresholding algorithm is employed, normalizing the cross-correlation function is imperative to nullify the impact of signal power:
\begin{equation}\label{eq:normalized_cross_correlation}
    CIR_{x,y}(\tau) = \frac{|R_{x,y}(\tau)|}{\sqrt{R_{x,x}(0) \times R_{y,y}(0)}} = \frac{|R_{x,y}(\tau)|}{\sqrt{P_x \times P_y}}
\end{equation}
where $P_x$ and $P_y$ denote the power of $x(t)$ and $y(t)$, respectively, and $|\cdot|$ represents the absolute value. The peak value of CIR is always $\leq 1$~\cite{briechle2001template}, irrespective of the signal power, indicating the similarity between the received signal $x(t)$ and the local template $y(t)$. Consequently, 4z employs dynamic thresholding~\cite{guvenc2005threshold} in the range from $0$ to $1$ to detect  ranging  packets based on channel conditions and SNR. Owing to the ultra-wide bandwidth of UWB signals, the CIR peak is exceedingly narrow, spanning around $2$ to $3$ nanoseconds. Therefore, the CIR peak can be utilized to precisely determine the timestamp of a UWB packet with adequate SNR, thereby serving as the foundation for accurate ranging as demonstrated in Eq.~\ref{eq:dstwr}.

\begin{figure}[b]
    \centering
    \includegraphics[width=1\linewidth]{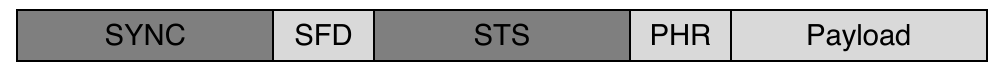}
    \vspace{-20pt}
    \caption{An example of UWB HRP packet configuration.}
    \label{fig:sp1_format}
    \vspace{-5pt}
\end{figure}

\subsection{UWB Secure Ranging}
\label{sec:packet_architecture}
Secure ranging refers to that the measured distance is always lower bounded by the actual physical distance even in the presence of an attacker. To achieve secure ranging, the 4z standard introduces a new field, i.e., scrambled timestamp sequence (STS), to prevent tampering with UWB packets at the physical layer. Figure~\ref{fig:sp1_format} shows an example of UWB HRP packet configuration with STS field. There are $5$ fields of a UWB HRP packet, namely synchronization (SYNC), start of frame delimiter (SFD), scrambled timestamp sequence (STS), physical layer header (PHD), and data payload. In this section, we mainly focus on the secure ranging which involves the SYNC and STS fields. We will discuss the other fields in Section~\ref{sec:physical_layer}.

\textbf{Synchronization and Packet Detection:} For accurate reception of data, the receiver needs to know when a packet begins. The SYNC field, also referred to as the preamble, is crafted to ascertain the presence of a UWB packet and to synchronize the receiver with the sender. This field encompasses a predefined ternary code in the alphabet $(1,0,-1)$, where $1$ denotes a positive pulse, $-1$ denotes a negative pulse, and $0$ denotes no pulse. The 4z standard specifies a limited number of preamble codes.
To enhance the SNR during NCC and the search for the CIR peak, a short preamble code is repeated multiple times. If the CIR peak surpasses a designated threshold, the receiver can confirm the presence of a UWB packet and synchronize itself based on the peak position. Specifically, UWB standards employ two distinct thresholds for presence detection and legitimate detection (synchronization). A lower threshold detects the presence of a UWB packet, even when the sender and receiver employ different preamble codes. In contrast, a higher threshold is utilized to verify the legitimacy of the packet, necessitating both sender and receiver to use the same preamble code. At each step, if there is no peak or the peak falls below the corresponding threshold, the receiver will drop the current packet and wait for the next packet. 

In the IEEE 802.15.4a~\cite{karapistoli2010overview} (4a) standard, the SYNC field serves a dual purpose, not only facilitating communication synchronization but also enabling ToF measurement and ranging. This is achieved through the CIR peak of the preamble, which provides an accurate timestamp for both sending and receiving the packet. However, the number of possible preamble codes is limited, and they are repetitively used in the SYNC field, making the ranging packet susceptible to tampering. This vulnerability creates an opportunity for attackers, allowing activities such as spoofing through the transmission of a forged legitimate copy of the packet in advance~\cite{poturalski2010cicada,dospaper}.

\textbf{Secure Ranging with STS Field:} To make the ranging results nontamperable, the 4z standard introduces a novel optional field, the Scrambled Timestamp Sequence (STS), within the UWB frame. Similar to the preamble, the STS functions as a sequence of pseudo-randomized pulses, distinct in that it is encrypted and does not repeat itself. The pseudo-random nature of the sequence eliminates periodicity, allowing the receiver to generate reliable, highly accurate, and artifact-free channel estimates. Decoding the STS requires the receiver to possess a local copy of the sequence, available before the start of reception. This is only feasible if both the transmitter and receiver know the keys and cryptographic scheme employed for STS generation. Importantly, the STS cannot substitute the preamble field since its correlation only functions after good synchronization.

Given the encrypted STS field in each packet, attackers are unable to prepare a copy of a ranging  packet in advance~\cite{leu2022ghost}. Consequently, a line-of-sight packet (directly from sender to receiver) always produces the first peak in the CIR. This occurs because light speed is the fastest, meaning any tampering with the packet will only result in a later peak in the CIR. Thus, the UWB ranging system can locate the first arrival peak to determine the packet timestamp, ensuring that the measured distance is always lower-bounded by the actual physical distance.
A recent work, the Ghost Peak~\cite{leu2022ghost} attack, injects random sequences into the STS field to create a fake peak before the actual first peak, attempting a distance reduction attack. However, the success rate of this attack is only $4\%$ as it relies solely on random sequences, making it unreliable in ensuring that the peak consistently precedes the actual first peak or surpasses the power threshold. Additionally, comparing the measured timestamps in the STS field with those from the SYNC field can help reject such a distance reduction attack~\cite{firaAccurateRanging, joo2023protecting}. The attack cannot generate the same time shift in both STS and SYNC fields due to the randomness, further reinforcing the reliability of HRP UWB ranging against distance reduction attacks by introducing the STS field~\cite{luo2023secure, joo2023protecting}, which cannot be forged.

\section{System and Threat Model}\label{sec:threat_model}

In this section, we provide an overview of the availability of COTS HRP UWB chips and their accessibility to the public. Subsequently, we outline our attack objectives and the capabilities possessed by potential adversaries.

\subsection{COTS HRP UWB Chips}
\label{sec:commercial_uwb_chips}

The market for UWB ranging is experiencing significant growth owing to its precise location sensing capabilities. Presently, commercially available COTS HRP UWB chips predominantly originate from three vendors: Apple, NXP, and Qorvo. 

\textbf{Apple} introduced the Apple U1 series chips in 2019, which have been integrated into various products such as the iPhone (since iPhone 11), HomePod mini, Apple Watch (since Series 6), and AirTag. 
\textbf{NXP} offers the Trimension series chips tailored for secure ranging and positioning, widely adopted by smartphone and car manufacturers including Samsung~\cite{ab40}, BMW~\cite{BWMUWB1}, Tesla~\cite{TeslaUWB}, and Volkswagen~\cite{Volkswagenuwb1}. 
These HRP UWB chips enjoy widespread adoption in consumer products. However, owing to security considerations, Apple and NXP do not disclose the configurations of the UWB packets for each product, resulting in non-interoperability between UWB ranging systems with differing packet configurations. On the other hand, \textbf{Qorvo} offers DW1000 and DW3000 series chips, providing a comprehensive set of low-level APIs that enable developers to configure the UWB physical layer flexibly. While the DW1000 chip only supports the 4a standard, the DW3000 series chips support the 4z standard. Consequently, the hardware prototype of \projname{} is built upon the DW3000 series chip, affording us the capability to sniff UWB packets using the provided APIs.

\subsection{Threat Model} 

The attacker's objective is to disrupt the commercial UWB ranging system by injecting malicious signals into the UWB communication channel, thereby impeding the update of distance data. Subsequently, the victim device may persistently rely on outdated distance data (e.g., in a PKES system) or fail to track a target (e.g., in iPhone and AirTag scenario). We establish the following assumptions for the attacker to achieve their objective:

\textbf{Capabilities of the Attacking Device: }
The attacking device must be developed using COTS UWB chips, allowing for the generation of ultra-wideband jamming packets at an affordable cost. Consequently, the attacking device can only manipulate UWB packets via the provided APIs of the COTS UWB chips. Additionally, the device must be compact and discreet, enabling it to be covertly positioned in hidden locations to launch attacks. Moreover, the attacking device should efficiently sniff UWB packets within a short timeframe to prepare for the attack, even in the case of encountering previously unseen commercial UWB ranging systems. 

\textbf{Knowledge of the Target UWB System: }
The adversary lacks access to confidential information exchanged between the targeted devices or the content of the unpredictable field (STS field) in HRP UWB, rendering it impossible to forge a ranging packet. Due to the unpredictability of the STS field, the attacker cannot preemptively send data packets to disrupt ranging functionality. While the attacker can freely receive and inject signals into the wireless communication channel, the structure of the victim's UWB packets remains undisclosed.

\textbf{Proximity Access to the Target UWB System:}
We assume that the adversary can physically approach the UWB system or discreetly install the attacking device in a hidden location, such as a car's undercarriage, beneath a table, or behind furniture. This proximity access can be facilitated when the car is temporarily parked, or the adversary prepares the attacking device in advance in a public space. Nevertheless, the attacker is unable to tamper with the hardware or software settings of the victim UWB system.

\begin{figure*}[t]
    \centering
    \begin{subfigure}[t]{0.33\linewidth}
        \centering  
        \includegraphics[height=2.8cm]{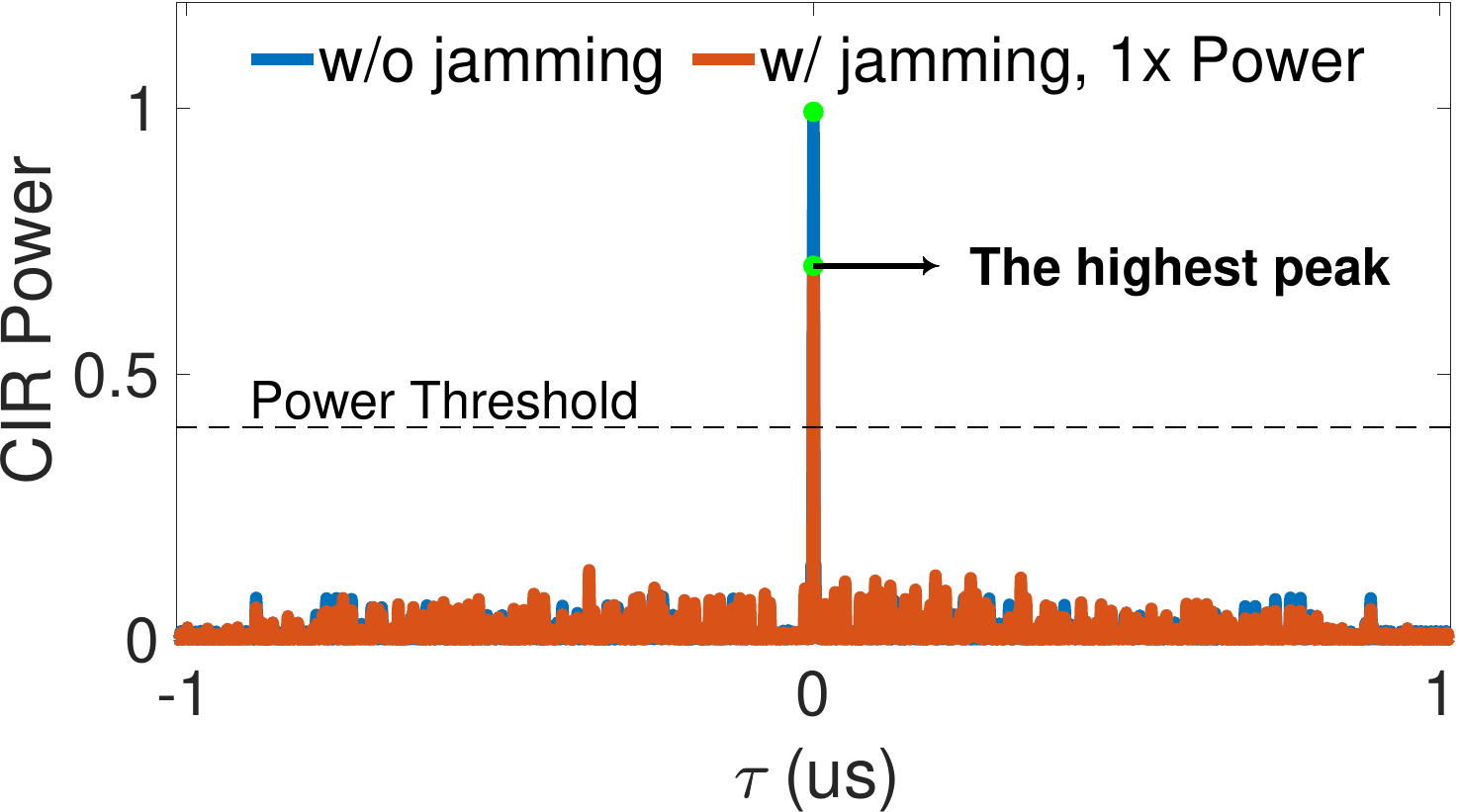}
        \caption{1x injecting power}\label{fig:jamming1power}
    \end{subfigure} 
    \begin{subfigure}[t]{0.33\linewidth}
        \centering
        \includegraphics[height=2.8cm]{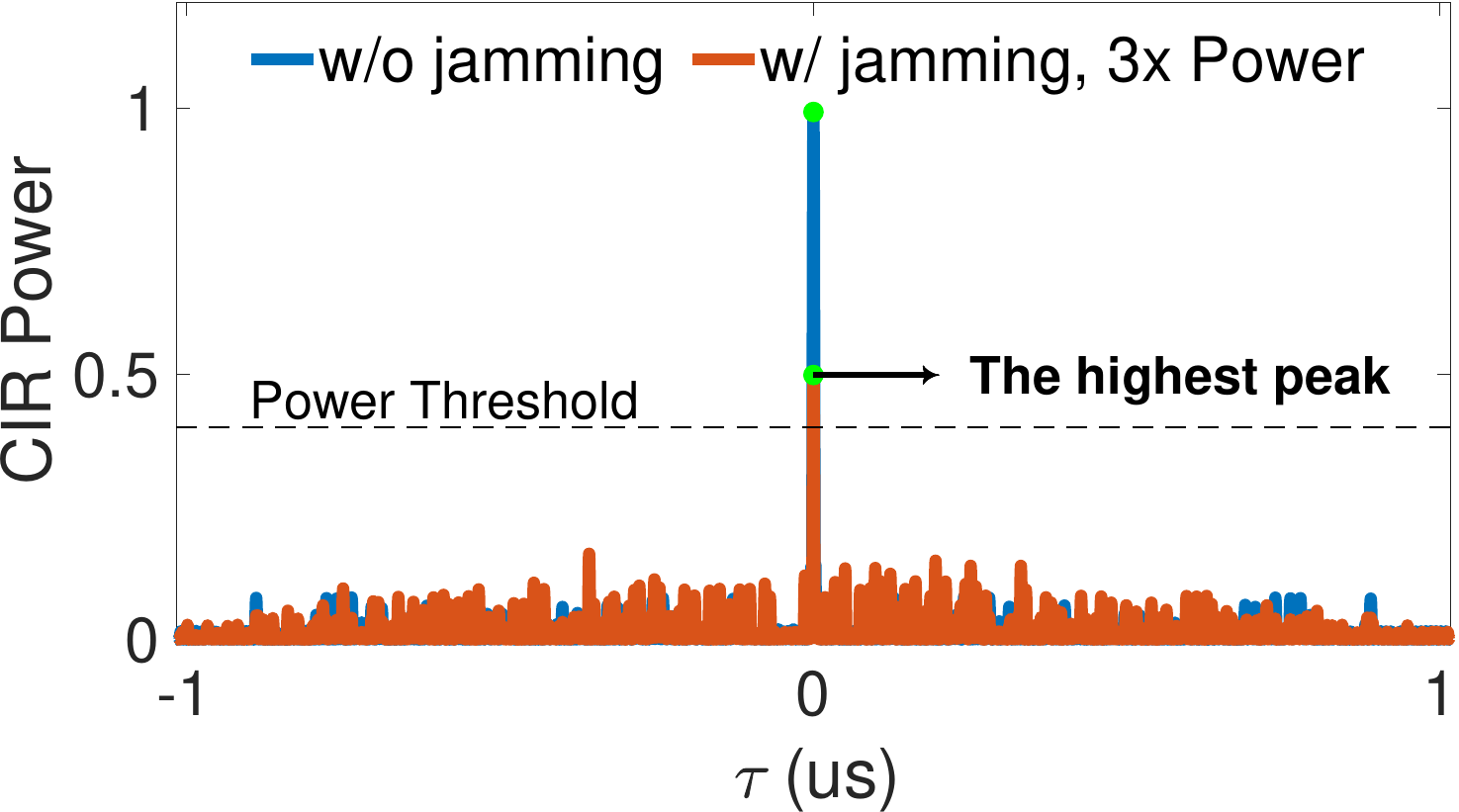}
        \caption{3x injecting power}\label{fig:jamming3power} 
    \end{subfigure} 
    \begin{subfigure}[t]{0.33\linewidth}  
        \centering  
        \includegraphics[height=2.8cm]{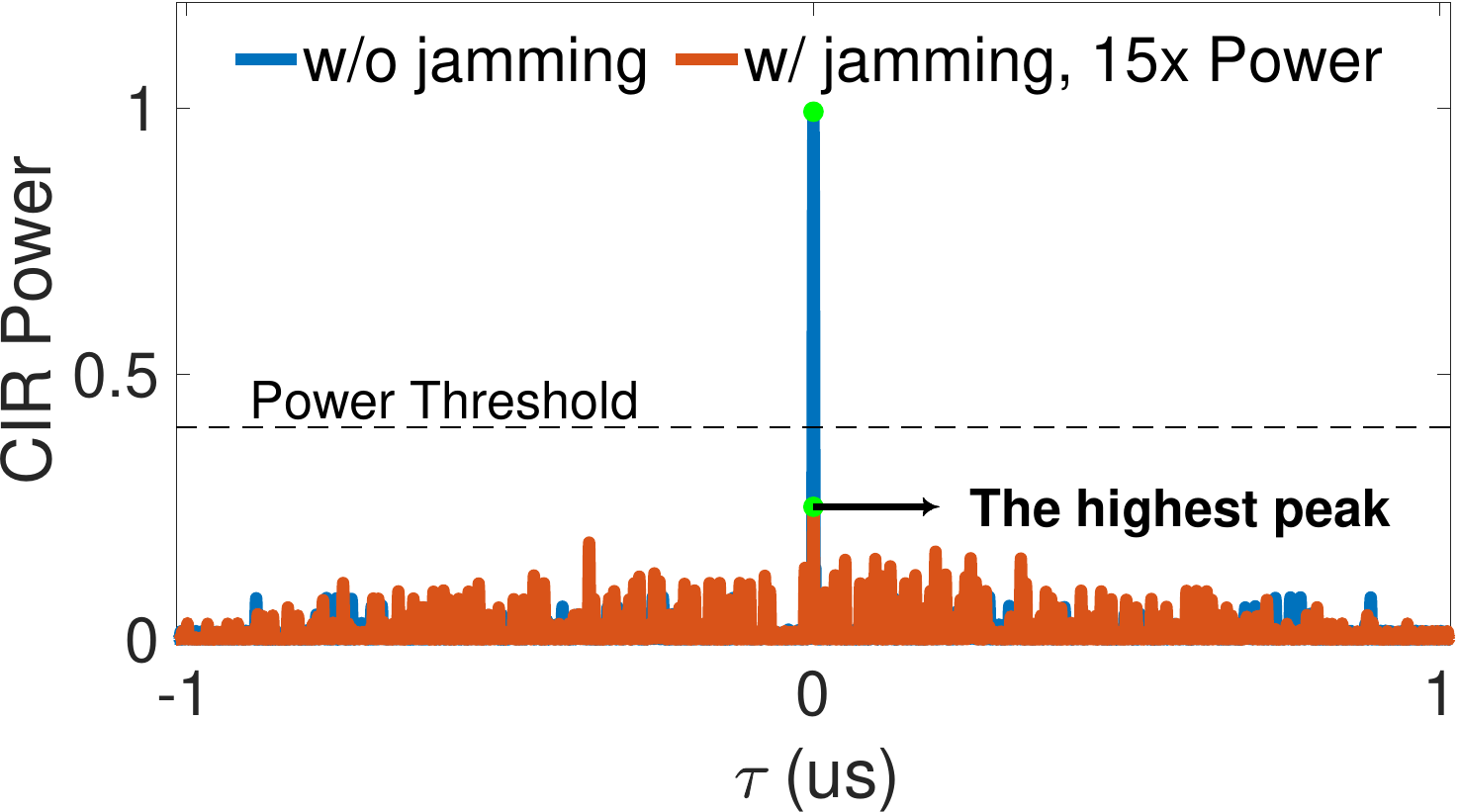}
        \caption{15x injecting power}\label{fig:jamming10power}
    \end{subfigure} 
    \vspace{-13pt}
    \caption{Injecting interference signal with larger power will significantly reduce the peak value of CIR estimated from the SYNC field, which may cause the ranging system to drop the current packet since the CIR peak is below the threshold.}
    \vspace{-10pt}
\end{figure*}

\section{Vulnerability Analysis}
\label{sec:core_idea}

In this section, we present our core idea to perform a jamming attack on the UWB ranging system and elaborate on the efficacy of injecting energy into the SYNC field to disrupt the ranging process. We also discuss the feasibility of implementing the jamming attack with COTS UWB chips.

\subsection{Leveraging NCC for jamming attack}
Using COTS UWB chips to generate the ultra-wideband jamming signals costs little, and the jamming signals cannot be simply removed by filtering. Nevertheless, due to the extremely short pulse and sparse modulation of UWB signals, it is difficult to conduct pulse-wise superposition for a jamming attack. UWB ranging systems adopt the NCC to merge the energy of multiple pulses of some fields (i.e., the SYNC and STS fields) to improve the SNR and estimate the CIR, which motivates us to leverage NCC to perform field-level jamming even if we cannot perfectly align the pulses of the jamming signal and the legitimate signal.
According to Eq.~\ref{eq:normalized_cross_correlation}, NCC computes CIR to compare the similarity regardless of the signal power, we can distort the received packet by superimposing attack packets with higher energy to the ranging packets. After that, the CIR peak will be decreased and drop below the detection threshold due to that the normalized power significantly increases.

For example, Figure~\ref{fig:jamming1power}-\ref{fig:jamming10power} compares the change of CIR peak with different power of the jamming signal, which clearly shows that the peak value is significantly decreased. The detection threshold is a dynamic value depending on the statistics of individual channel realizations and the relative energy of the received signal~\cite{guvenc2005threshold}. We take $0.4$ as an example here. In this case, the UWB ranging system may drop the packet once the peak CIR is below the threshold and not update the ranging result. By continuously injecting the jamming signal for each ranging process, the ranging system can be disrupted, and the recorded distance data may be incorrect or the system may report no signal.




\subsection{Theoretical analysis}\label{sec:jamming_principal}
\begin{figure}[t]
    \centering
    \includegraphics[width=0.8\linewidth]{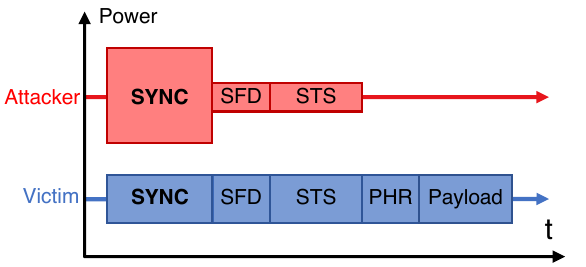}
    \vspace{-5pt}
    \caption{Attack packet with amplified power in the SYNC field. The attack packet and legitimate ranging packet should be aligned in time for jamming attack.}\label{fig:jamming}
    \vspace{-15pt}
\end{figure}

Suppose the jamming packet is perfectly synchronized with the ranging packet, i.e. they arrive at the receiver at precisely the same time, as shown in Figure~\ref{fig:jamming}. The energy of the SYNC field of a jamming packet is amplified, while the energy of the other fields remains unchanged. We first analyze the impact of injecting energy to the SYNC field on the peak value of estimated CIR, and then discuss why attacking the SYNC field can maximize the efficiency in Sec.~\ref{sec:why_sync}. 

\begin{figure*}
    \centering
    \includegraphics[width=0.9\linewidth]{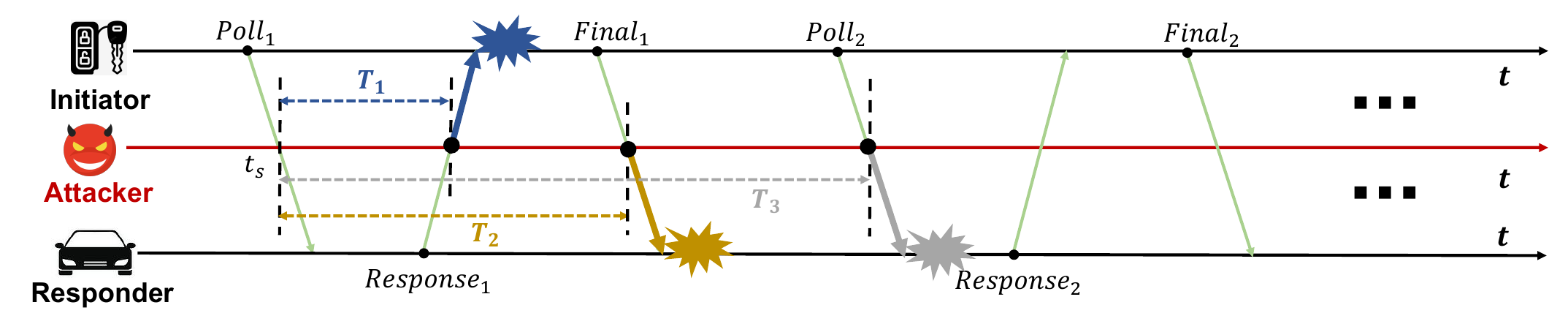}
    \vspace{-13pt}
    \caption{After listening to several ranging sessions, \projname{} can (i) determine the start time of ranging and (ii) the time delay between each packet, i.e., $T_1$, $T_2$, and $T_3$. Since these time intervals are always fixed for a specific commercial product, \projname{} can predict the arrival time of the following packets and launch the attack at the correct timing.}
    \label{fig:attack_scenarios}
    \vspace{-10pt}
\end{figure*}

Let $x(t)$ be the local template of the SYNC field, $x'(t)$ be the signal traveling from the sender to the receiver, and $y(t)$ be the jamming signal with enlarged power. The received signal at the receiver side is the superposition of the jamming signal and the legitimate signal, i.e., $x'(t) + y(t)$. Assume we can find the peak CIR without jamming at time $\tau^*$, i.e., $CIR_{x,x'}(\tau^*)$. \zyz{Then, by substituting the received signal $x'(t) + y(t)$ and local template $x(t)$ to Eq.~\ref{eq:normalized_cross_correlation}, the peak value of the CIR after the jamming attack becomes:}
\begin{equation}\label{eq:5}
    CIR_{x,x'+y}(\tau^*) = \frac{|R_{x,x'+y}(\tau^*)|}{\sqrt{R_{x,x}(0)R_{x'+y,x'+y}(0)}}
\end{equation}
\zyz{
Given that cross-correlation is a linear operation, we have $R_{x,x'+y}(\tau^*) = R_{x,x'}(\tau^*)+R_{x,y}(\tau^*)$. We assume the jamming packets are not correlated with the ranging packets, i.e., $R_{x,y}(t)\approx 0$ at any given time $t$. This can easily be achieved by using different preamble codes for the jamming packets and the ranging packets. Then we have:
\begin{equation}\label{eq:5_remove_y}
    CIR_{x,x'+y}(\tau^*) = \frac{|R_{x,x'}(\tau^*)|}{\sqrt{R_{x,x}(0)R_{x'+y,x'+y}(0)}}
\end{equation}
Recall that $CIR_{x,x'}(\tau^*) = \frac{|R_{x,x'}(\tau^*)|}{\sqrt{R_{x,x}(0) \times R_{x',x'}(0)}}$ represents the CIR peak of the SYNC field without jamming, then we can rewrite Eq.~\ref{eq:5_remove_y} as:} 
\begin{align}\label{eq:attack_model}
    CIR_{x,x'+y}(\tau^*) &= \sqrt{\frac{R_{x',x'}(0)}{R_{x'+y,x'+y}(0)}}CIR_{x,x'}(\tau^*)\notag \\
    &=\sqrt{\frac{R_{x',x'}(0)}{R_{x',x'}(0)+R_{y,y}(0)+2R_{x',y}(0)}}CIR_{x,x'}(\tau^*)\notag\\
    &=\sqrt{\frac{P_{x'}}{P_{x'}+P_y}}CIR_{x,x'}(\tau^*)
\end{align}
where $P_{x'} = R_{x',x'}(0)$ is the power of the victim signal, $P_y = R_{y,y}(0)$ is the power of the jamming packets, and $R_{x',y}(0)$ is the cross-correlation between the victim signal and the jamming packets, which is, similarly, close to zero since $x'$ and $y$ are not correlated. Clearly, the CIR peak of the SYNC field after jamming will be decreased by a factor of $\sqrt{\frac{P_{x'}}{P_{x'}+P_y}}$, which is smaller than $1$. Therefore, once the power of jamming packets is large enough to make the CIR peak fall below the threshold (e.g., $0.4$), the UWB ranging system will drop the packet and fail to update the ranging result. 

We test the impact of jamming packets in simulation using the official simulator~\cite{matlabuwb} in MATLAB. To be specific, we choose two preamble codes from the 4z standard, e.g., code index $9$ for the ranging packet and code index $12$ for the jamming packet, and then generate the corresponding physical layer signals and amplify the power of the jamming packet by $1\times$, $3\times$, and $15\times$, respectively. Next, we evaluate the change of CIR peak after jamming with different signal power. The results are shown in Figure~\ref{fig:jamming1power}-\ref{fig:jamming10power}. We have the following two observations: (i) The noise floor does not change with the increased power of the jamming packets, due to the use of normalization. This implies that the NCC only compares the similarity of two signals regardless of the signal power. (ii) The peak CIR of the SYNC field is decreased with the increased power of the jamming packets, by a factor of around $0.7$, $0.5$, and $0.25$, respectively, which is consistent with our model in Eq.~\ref{eq:attack_model}. This is not a surprise since the jamming packets with larger power will decrease the similarity between the superposed signal and the local template.

Uncorrelated jamming packets may significantly decrease the CIR peak, while the correlated noise (multipath effect, especially when there is no Line-of-Sight signal) may have limited impact. Correlated multipath noise will lead to multiple peaks in the CIR spectrum and also decrease the value of all CIR peaks. However, we can apply iterative subtraction to eliminate the highest peak until find the first peak that is above the threshold to denote the shortest path, as specified in the 4a standard.

\subsection{Reasons for Attacking the SYNC Field}\label{sec:why_sync}
The working principle of our attack works for both the SYNC field and STS field since both of them rely on NCC. However, we choose to attack the SYNC field instead of the STS field due to $3$ reasons: (i) Our experiments indicate that attacking the SYNC field alone can already yield a high success rate. (ii) The SYNC field adopts ternary code, while the STS field adopts binary code. This means that the SYNC field has lower average power compared to the STS field since there is no pulse if the code is $0$. According to Eq.~\ref{eq:attack_model}, the higher power of the victim signal requires the higher power of the jamming signal to conduct the attack, making it harder to mount a jamming attack. (iii) Although the time duration of the SYNC field can be similar to that of the STS field in some cases, the STS field has a longer code without repetition, while the SYNC field uses a much shorter code with tens of repetitions. This makes the STS field, compared to the SYNC field, have better correlation properties, and thus we have to inject more energy to decrease the CIR peak. 

Due to the sparse modulation of UWB pulses, it is difficult to conduct pulse-wise superposition to jam the fields carrying data streams such as PHD and payload. Moreover, PHD and payload jamming can be easily detected by using the transmission statistics~\cite{mykytyn2021jamming}. For the SFD field, we cannot control its power with the given API, while the output power of the other fields can be tuned. 
The attack packet in Figure~\ref{fig:jamming} does not contain the PHD and payload fields to better determine the packet length, which will be discussed in Sec.~\ref{sec:attack_delay}.

\subsection{Methods for Jamming Packets in Time}
The duty cycle of impulse radio UWB is very low and even less than $1\%$ in some cases~\cite{qian2022overview}. It means that the attacker has to know the right timing to inject the jamming packets to superpose with the ranging packet at the receiver side. Recall that we use the COTS UWB chips to generate the ultra-wideband jamming packets. The next question is that how do we inject the jamming packets in time with the well-encapsulated chips and the given APIs.

\textbf{When does the ranging start?} 
The first question to perform a jamming attack is to determine when the ranging process starts. Our reactive jamming should be able to continuously listen to the communication and react after receiving the first packet (record the receiving timestamp of the first packet). However, for well-encapsulated UWB chips, the sender and receiver need to maintain consistency in the entire UWB physical layer structure to read the timestamp. In other words, we have to sniff the complete physical layer structure of the UWB packets, because it is not publicly available for commercial products. After that, we can know the exact received timestamp of each ranging packet.

\textbf{Can we predict the time of the next packet?}
After recording the timestamp of the packets in several ranging processes and computing the time delay between each packet, we can predict the arrival time of the following packets and launch the attack at the right time. This is due to the fact that the duration between each packet in one ranging process is fixed, and the duration between two ranging processes is also fixed, as specified in the UWB standards. This observation is applicable for standards including 4a, 4z, and even the upcoming 4ab, which provides us an opportunity to perform a jamming attack at the right time.

Specifically, in a UWB ranging system, as shown in Figure~\ref{fig:attack_scenarios}, to complete one ranging process, the initiator and the responder will transmit $2$ to $3$ messages, corresponding to the SS-TWR and DS-TWR, respectively. Therefore, \projname{} is capable of attacking the following packets if (i) the start timestamp $t_s$ is known and (ii) the time delay $T_1$ (between poll and response), $T_2$ (between poll and final), and $T_3$ (between two consecutive polls) are known, where $T_3$ should be much larger than $T_1$ and $T_2$. Since the ToF is in the order of nanoseconds and \projname{} aims at field-level superposition, we can ignore the impact of ToF. Actually, it is not necessary to jam all the packets, since only one packet loss will disrupt the ranging process. Therefore, we jam the response message, since it works for both SS-TWR and DS-TWR. The poll message can help us to correct the predicted time for a long-term attack.

\section{Attack Implementation}
\label{sec:design}
\begin{figure*}[htbp]
    \centering
    \includegraphics[width=\linewidth]{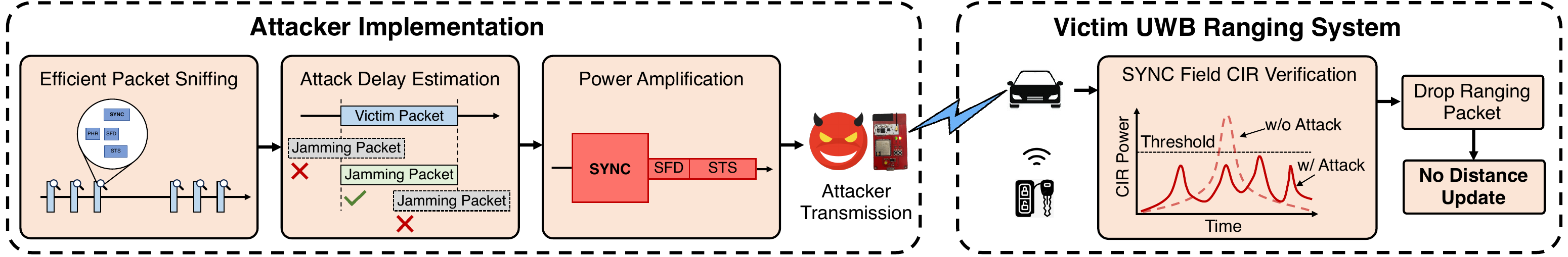}
    \vspace{-10pt}
    \caption{\projname{} consists of three key modules to launch the jamming attack, including (i) packet sniffing, (ii) attack delay estimation, and (iii) power amplification. The jamming packet arrives at the victim device at the same time as the ranging packet. After that, the CIR peak in the SYNC field will drop below the power threshold, then the receiver drops the ranging packet. If we can successfully attack every ranging process (cause packet drop), there will be no ranging distance update at the victim device.}\label{fig:overview}
    \vspace{-10pt}
\end{figure*}

To launch an imperceptible and reliable jamming attack against the UWB ranging system, \projname{} incorporates three key modules, illustrated in Figure~\ref{fig:overview}. First, when dealing with previously unseen devices, \projname{} must rapidly sniff the packet configuration, enabling the extraction of timestamps from received packets and facilitating the launch of a jamming attack within a short preparation time. Second, accurate estimation of the attack delay is crucial for \projname{}, considering that the measured time interval between two consecutive packets cannot be directly applied. Third, to generate the attack packet, \projname{} needs to select a different preamble code and amplify the power of the SYNC field to ensure the effectiveness of the attack.
The targeted UWB ranging system will drop the ranging packet during its verification of the CIR peak in the SYNC field because it detects that the peak falls below the power threshold under our attack. Successfully disrupting each ranging process results in the distance measurement not being updated, leading to a complete failure of the victim device's UWB ranging functionality.
In this section, we delve into the details of our attack implementation.


\subsection{Packet Sniffing}\label{sec:physical_layer}
The UWB physical layer consists of multiple adjustable parameters for the first $4$ fields shown in Figure~\ref{fig:sp1_format}, while the last field (i.e., payload) only contains transmitted data. To receive the complete UWB packet, the attacker needs to sniff the parameters of the first $4$ fields. However, the exhaustive search and manual tuning of these parameters is a tedious and time-consuming task. Therefore, we propose an efficient packet sniffing algorithm to make \projname{} applicable for unseen UWB devices.





\textbf{Possible parameters for packet sniffing.}
Each field contains multiple adjustable parameters and each parameter may have many options. For example, there are $4$ parameters to determine the SYNC field, i.e., the UWB channel number (\textit{chan}), the preamble length (\textit{txPreambLength}), the preamble code (\textit{tx/rxCode}), and the preamble chunk size (\textit{rxPAC}). For the STS field, we have to determine the STS mode (\textit{stsMode}) and the STS length (\textit{stsLength}). The Start-of-Frame Delimiter (SFD) field indicates the delimitation of the frame, which is mainly determined by the type of SFD field (\textit{sfdType}). The physical header (PHD) field is used to provide the receiver with detailed information about the payload, which comprises the data rate (\textit{dataRate}), the PHD mode (\textit{phdMode}), the PHD rate (\textit{phdRate}), and PDOA mode (\textit{pdoaMode}). For the details about each parameter, please refer to DWM3000 API Guide~\cite{DWM3000API}.
The number of all possible combinations of these parameters exceeds $100k$. Suppose each configuration requires $5s$ to test, it would take over $100$ hours to try all possibilities. Therefore, it is crucial to find an efficient way to sniff the packet structure automatically and efficiently, which can be then applied to unseen UWB devices (without knowing any information about the victim UWB ranging system ahead of time).

\textbf{Efficient Packet Sniffing.}
Our key observation to achieve efficient packet sniffing is that the receiver checks the configuration of the packet field by field and has different responses (i.e., error codes) for each field.
This property makes it possible to sniff the packet structure efficiently by only changing the parameters of one field at a time, which largely reduces the potential searching space.
To be specific, we have $4$ steps to sniff the packet structure.
First, our attacking device continuously listens to the wireless channel and waits for the UWB packets, by iteratively tuning the two parameters: the channel number (\textit{chan}) and the chunk size (\textit{rxPAC}). If these two parameters are correct, the packet is detected and there will be an error code indicating that the SYNC field is incorrect.
Second, we test the combinations of the preamble code (\textit{tx/rxCode}) and the preamble length (\textit{txPreambLength}) iteratively until the error code disappears.
Third, similarly, we test the type of SFD field (\textit{sfdType}) until its error code disappears.
Lastly, since the STS field and PHD field share the same error code, we need to determine all their parameters simultaneously. Luckily, the parameters of the PHD field are related, thus they do not have many combinations.
This approach greatly reduces the search space from over $100k$ to around $300$.
Note that, we can only test one configuration for each packet since the UWB chip will drop this packet if the configuration is incorrect.

\subsection{Attack Delay Estimation}\label{sec:attack_delay}

Though we can sniff the physical layer structure of the UWB packets and record the timestamp of each packet received, we cannot directly compute the attack delay (i.e., $T_1$, $T_2$, and $T_3$ as shown in Figure~\ref{fig:attack_scenarios}) by subtracting the measured timestamp. The reason is, whenever the attacking device receives a packet, this means that the packet has already been transmitted and exists in the wireless channel. However, the attacker needs to transmit the jamming packets such that it arrives at the same time as the ranging packets at the victim device. The receiving and transmitting processes require some time themselves. Therefore, we need to tune down the measured time to compute the actual attack delay. To be specific, we have:
\begin{equation}
    T_{attack} = T_{measure} - T_{chunk} - T_{packet} - T_{\Delta}
\end{equation}
where $T_{measure}$ is the time difference of two measured timestamps, $T_{chunk}$ is the receiving time for preamble detection at the receiver side (i.e., $rxPAC$), $T_{packet}$ is the time duration of a transmitted packet (the preparing time at the transmitter side), and $T_{\Delta}$ is the transmission delay. $T_{chunk}$ and $T_{packet}$ can be sniffed from the physical layer structure, as shown in Figure~\ref{fig:jamming}. Note that, our attack packet only contains the first three fields (i.e., SYNC, STS, and SFD), a packet format allowed by the COTS chips, because the length of this packet is more predictable. $T_{\Delta}$ is almost the same for different devices and configurations (e.g., AirTag, iPhone, and DWM3000EVB with various configurations), which is around $20\mu s$ in average and standard deviation $2.74\mu s$ in our experiments.

\begin{figure}
    \centering
    \includegraphics[width=0.95\linewidth]{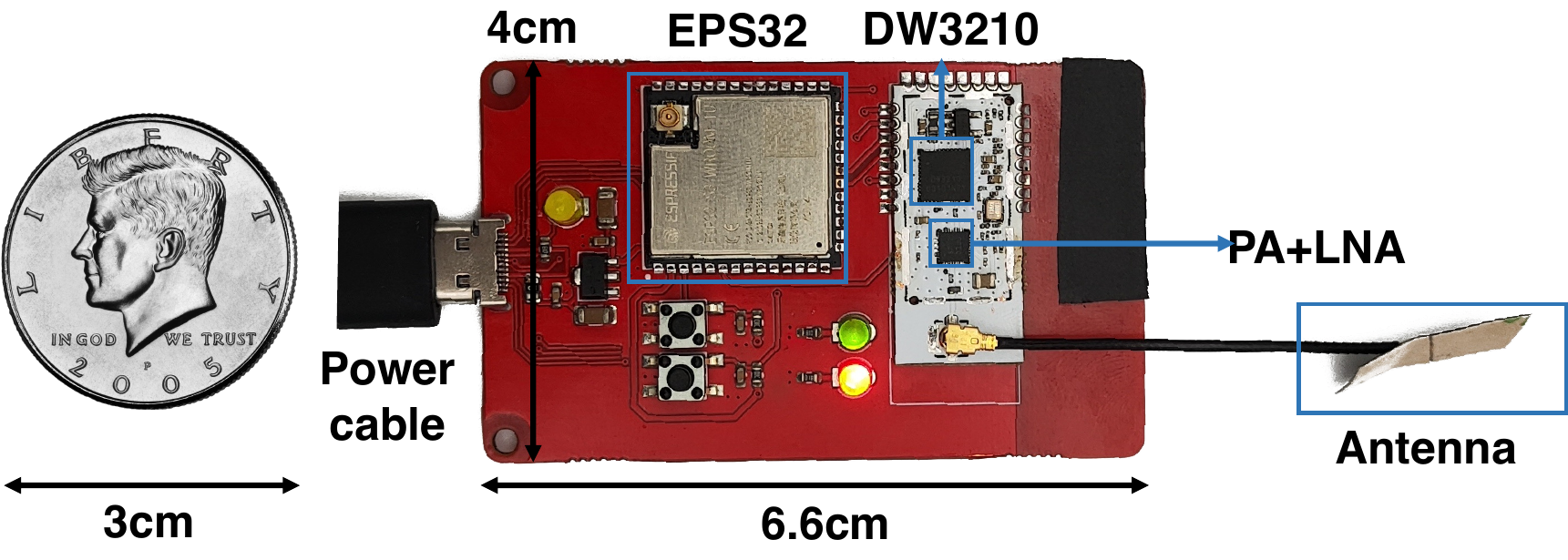}
    \vspace{-10pt}
    \caption{Hardware prototype of \projname{}.}
    \label{fig:uwb-hw-arch}
    \vspace{-15pt}
\end{figure}

\subsection{Hardware Design}
There are two basic requirements for the hardware design of \projname{}: (i) it should be developed based on COTS UWB chips for low-cost ultra-wideband jamming and easy deployment, and (ii) it should have enough output power to interfere with legitimate UWB signals.

The hardware design is shown in Figure~\ref{fig:uwb-hw-arch}.
We choose Qorvo DW3210~\cite{dw3210} as the UWB chip, which supports the 4z standard and provides a rich set of low-level APIs. Then we use a microcontroller (ESP32~\cite{babiuch2020smart}) to communicate with the DW3210 chip following the guidelines of \cite{DWM3000API,dw3000andesp32}. This allows us to sniff packets and launch attacks automatically.
After that, we add an amplifier module (TQP3M9035~\cite{9035}) at the antenna port, consisting of a power amplifier (PA) to boost the output power, and a switching circuit for controlling the transmit and receive modes of the antenna port. 
The output power of the DW3210 is $-41.3dBm/MHz$ according to its datasheet~\cite{dw3210}, which is also the maximum power of UWB signals as specified by regulations~\cite{krebesz2017use}. 
In contrast, the maximum output power after adding the power amplifier reaches $-23dBm/MHz$, introducing over $18dB$ signal gain and making it more flexible for jamming attacks.

\section{Experimental Evaluation}
\label{sec:eval}
We evaluate the performance of \projname{} with UWB modules (Sec.~\ref{sec:performance}) and commercial products (Sec.~\ref{sec:case_study}).

\subsection{Attack Effectiveness}\label{sec:performance}
The efficacy of our system design is evaluated with the following experimental setup.

\textbf{Victim devices:}
We adopt two commercial UWB modules~\cite{taobao} (developed based on DW3210) as the victim initiator and responder, because they are fully controllable, making it convenient for us to verify the performance of \projname{} at different settings, such as different the packet structure and retrieve the exact time intervals between consecutive packets. We set the power of the victim devices to strictly below the maximum power allowed by the FCC regulations.

\textbf{System configuration:}
The UWB ranging system of the victim devices is configured as follows. We use DS-TWR ranging and HRP mode. The length of the SYNC field and STS field is set to be $64\mu s$. Other parameters of packet structures are selected randomly for each test. The victim devices start a new ranging session every $167 ms$. The time interval between the poll packet and the response packet is $800\mu s$, and that between the poll packet and the final packet is $1600\mu s$.


\textbf{Metrics: }
We conduct \projname{} attack on the response packet (from the responder to the initiator).
We repeat ranging sessions for $30s$ for each test and count the number of received response packets ($N_r$) and the number of sending poll packets ($N_p$) at the initiator. Then we calculate the success rate $1-N_r/N_p$ as our metric. If the success rate reaches $100\%$, it means that we can block all the ranging sessions and completely fail the ranging process of the victim devices.

\subsubsection{Efficiency of Packet Sniffing}
We first examine the efficiency of our packet sniffing algorithm. The initiator is positioned in the middle and $1m$ away from both the responder and the attack device. We configure $30$ random physical layer structures for UWB ranging between the two victim devices. Subsequently, we employ \projname{} to sniff the packet structure and record the average time taken for sniffing each field. The sniffing time for each field is documented based on the responses obtained from \projname{}. The results are detailed in Table~\ref{tab:sniffing_time}.
As evident from the results, the average sniffing time for each packet configuration is approximately $22.46s$. Notably, sniffing the SYNC field consumes the most time, constituting around $82\%$ of the total time. This is attributed to the SYNC field's greater number of adjustable parameters. Note that, during sniffing, \projname{} can only test one potential packet configuration for each received packet. Therefore, the sniffing time of \projname{} is contingent upon the ranging frequency of the victim devices. Considering that victim devices engage in ranging $6$ times per second, \projname{} requires $135$ ranging sessions to sniff the packet structure. 

\begin{table}[b]
    \setlength{\textwidth}{3pt}
    \centering
    \caption{Average sniffing time for different fields.}
    \begin{tabular}{cccccccc}
            \hline
            \textbf{SYNC} &\textbf{SFD }& \textbf{STS and PHD} &\textbf{Total Time}\\
             \hline
            17.94s &  0.83s & 3.69s& 22.46s\\
               \hline
    \end{tabular}
    \label{tab:sniffing_time}
\end{table}

\begin{figure}
    \centering
    \includegraphics[width=0.7\linewidth]{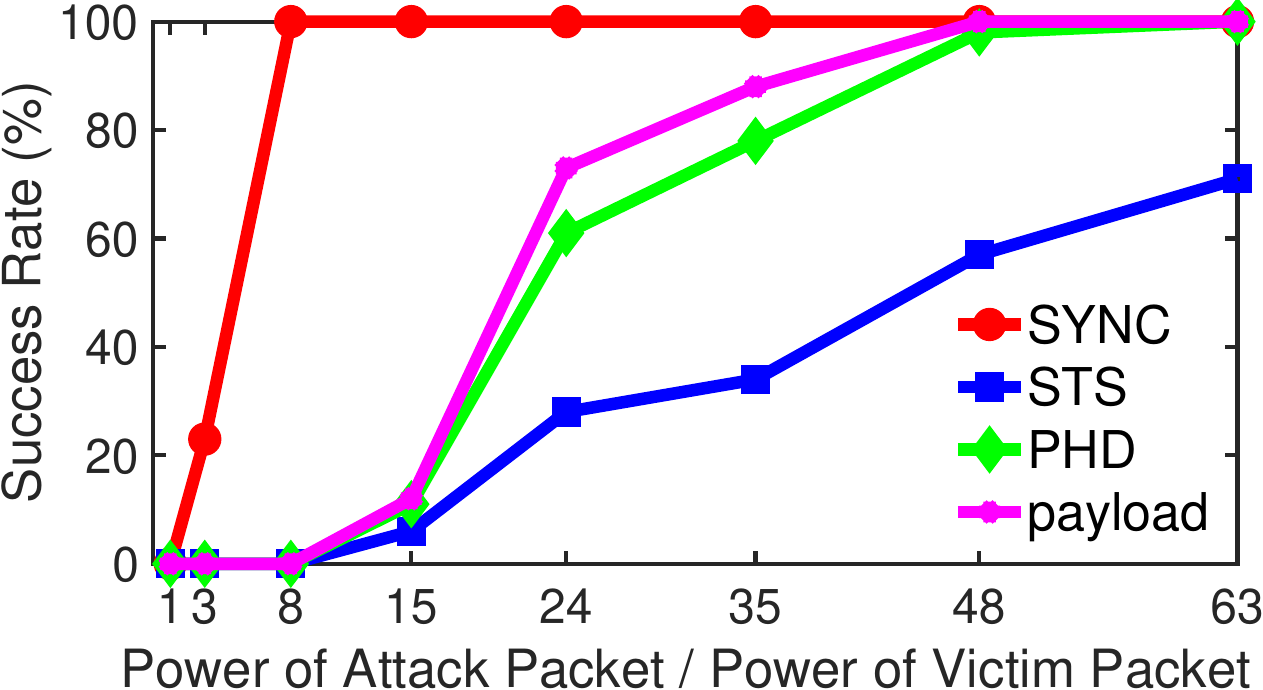}
    \vspace{-10pt}
    \caption{Impact of attacking different fields.}
    \label{fig:field}
    \vspace{-15pt}
\end{figure}


\subsubsection{Impact of Target Field}

We assess the impact of attacking different fields on the success rate under various power gains (i.e., power of the attack packet divided by power of the victim packet), given the accurate attack delay. The distance between the victim devices and that between \projname{} and the initiator remains consistent (i.e., $1m$). Since all testing devices are developed based on DW3210, we assume the power of the attack packet is equivalent to that of the victim packet. Consequently, we can adjust the output power of \projname{} to control the power gain. We evaluate the attack performance across four different fields: SYNC, STS, PHD, and payload. SFD is disregarded due to the absence of APIs to individually tune its power. The results are depicted in Figure~\ref{fig:field}. Notably, the success rate of attacking the SYNC field converges to $100\%$ when the $8\times$ power gains, whereas the PHD and payload fields require $48\times$ power gain even with shorter field lengths. This discrepancy primarily arises because PHD and payload fields carry data streams and do not utilize NCC for channel estimation. Attacks on PHD and payload fields also result in packet loss due to cyclic redundancy check (CRC). However, we refrain from attacking PHD and payload fields as they require higher power compared to the SYNC field and are more susceptible to detection through data stream statistics. Moreover, the success rate of attacking the STS field does not reach $100\%$ even with the highest power gain of $63$ and the same field length as the SYNC field, mainly due to the modulation scheme and non-repetitive sequence of the STS field, as discussed in Sec.~\ref{sec:jamming_principal}.

\subsubsection{Impact of Attack Delay}
\begin{figure}
    \centering
    \includegraphics[width=0.73\linewidth]{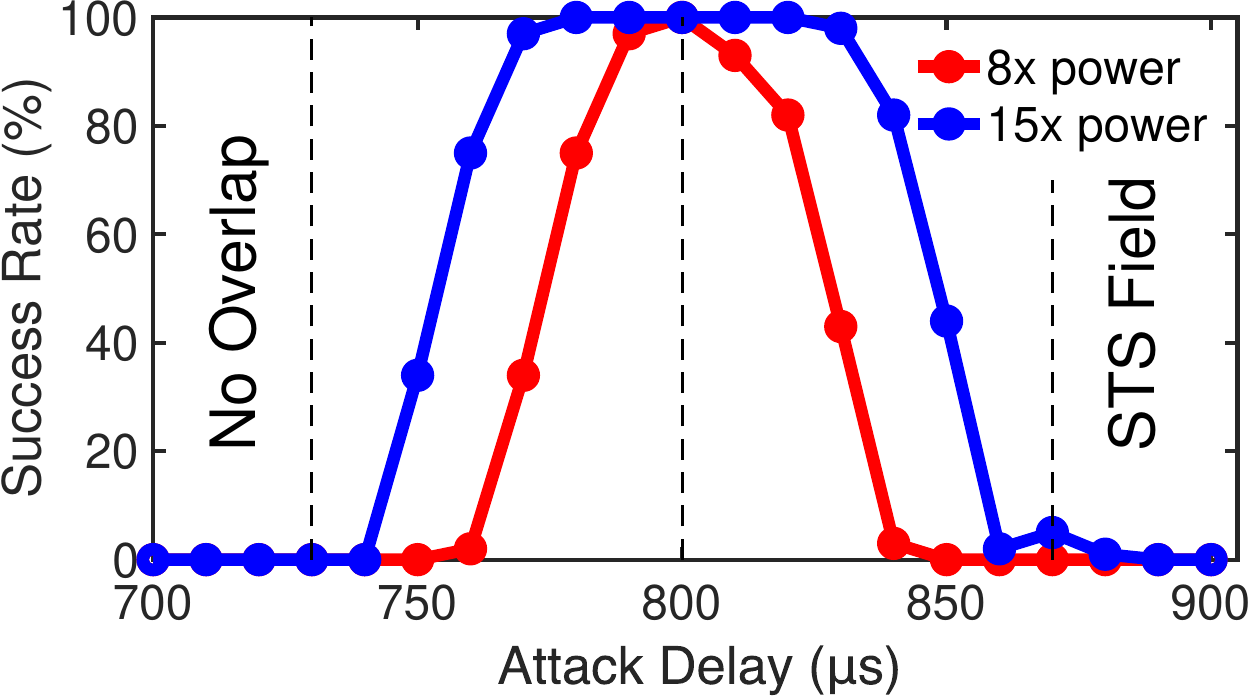}
    \vspace{-10pt}
    \caption{Success rate under varying attack delay. \zyz{Increasing the power enhances the tolerance for attack delay errors, leading to an almost 100\% success rate.} }
    \label{fig:attack_delay}
\end{figure}

Given that the length of the SYNC field is $64\mu s$ and the accurate attack delay is $800\mu s$, we vary the attack delay of \projname{} from $700\mu s$ to $900\mu s$ and evaluate the success rate under different delay errors. Three distinct conditions arise within this delay range: no field overlap (i.e., less than $736\mu s$), overlap with SYNC (i.e., $736\mu s$ to $864\mu s$), and overlap with STS (i.e., greater than $864\mu s$). We set the power gain to $8$ and $15$. The results are presented in Figure~\ref{fig:attack_delay}. Initially, with $8\times$ power gain, the success rate only reaches $100\%$ when the SYNC fields are perfectly overlapped while increasing the power gain to $15$ enhances the tolerance of delay errors in practical attacks. \zyz{This implies that the attack delay estimation can be relatively coarse (within the length of the SYNC field) in practice if the power of the jamming signal is sufficiently high, which can be easily achieved by reading the packet timestamp. Also, a more precise attack delay estimation allows for a longer attacking distance at the same power level.}
Furthermore, when the SYNC field of the attack packet overlaps with the STS field of the victim packet at around $870\mu s$ with $15\times$ power gain, it achieves approximately $6\%$ success rate but rapidly diminishes to $0$. This trend aligns with the findings in Figure~\ref{fig:field}, further indicating that attacking the SYNC field can be considerably more robust to attack delay errors compared with attacking the STS field, due to the power efficiency.



\begin{figure}
    \centering
    \includegraphics[width=0.7\linewidth]{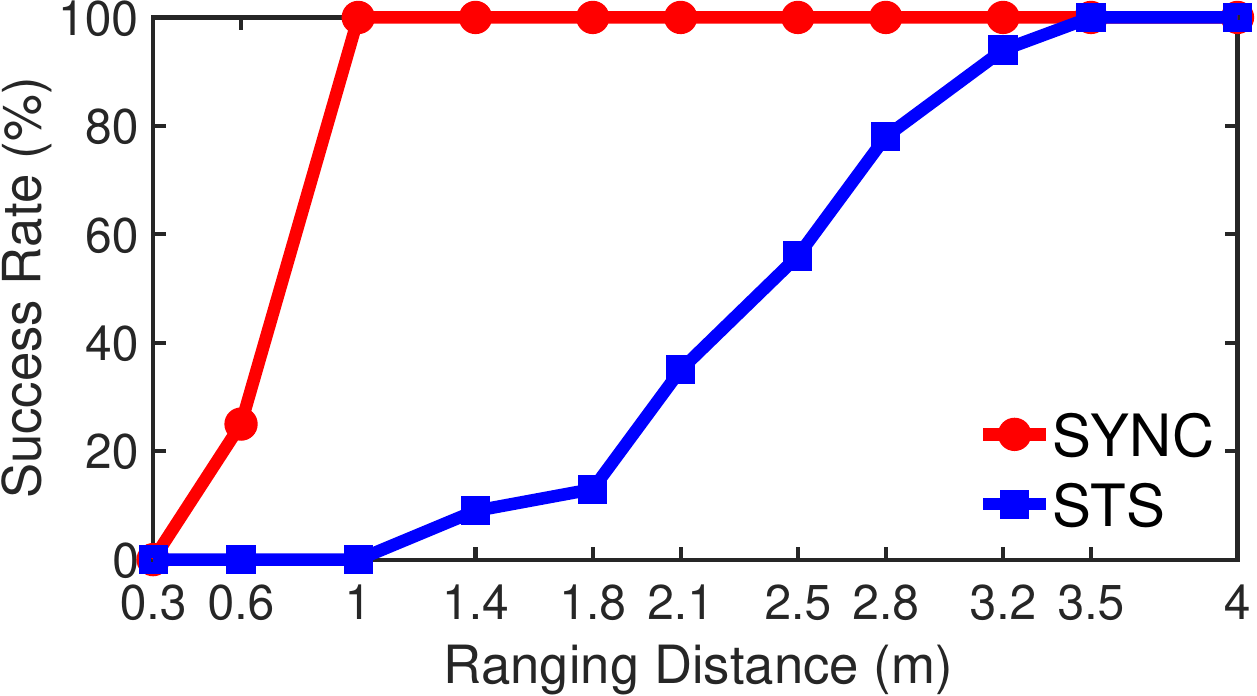}
    \vspace{-10pt}
    \caption{Impact of ranging distance (between the initiator and responder).}
    \label{fig:distance}
    \vspace{-15pt}
\end{figure}

\subsubsection{Impact of Ranging Distance}

Next, we explore the impact of ranging distance (i.e., the distance between the initiator and the responder) on the success rate. \projname{} is positioned $1m$ away from the initiator with $8\times$ power gain. We document the success rate at the varying ranging distance from $0.3m$ to $4m$ (altering the responder's position). The results are displayed in Figure~\ref{fig:distance}. We can make the following two observations: (i) The success rate for attacking the STS field reaches $100\%$ at around $3.5m$, whereas that of attacking the SYNC field is only $1m$. This implies that in practical attacks, the UWB ranging system needs to approach much closer to restore ranging functionality if the attacker targets the SYNC field. (ii) The ToF at various ranging distances does not affect the attack performance, even the $8\times$ power gain is just sufficient to achieve $100\%$ success rate with accurate attack delay.

\begin{figure*}[t]
    \centering
    \begin{subfigure}[b]{0.33\linewidth}
        \centering
        \includegraphics[width=1\textwidth]{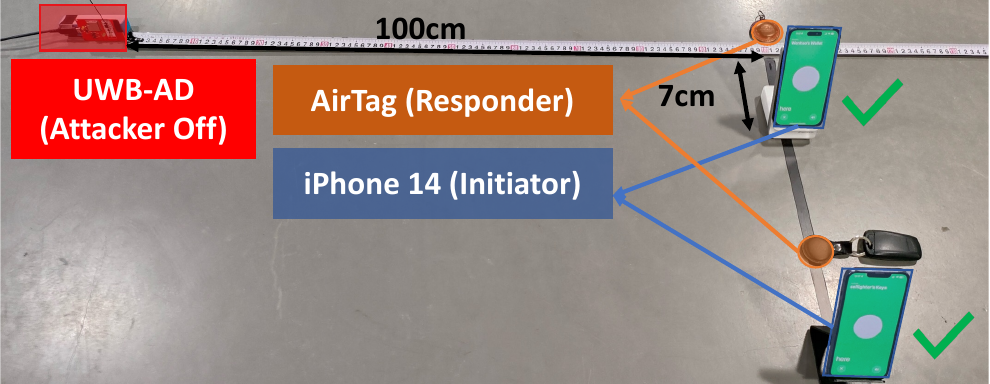}
        \caption{\zyz{Setup of attacking iPhone-AirTag pairs.}}
        \label{fig:airtag1}
        \vspace{1pt}
    \end{subfigure}
    \begin{subfigure}[b]{0.33\linewidth}
        \centering
         \includegraphics[width=1\textwidth]{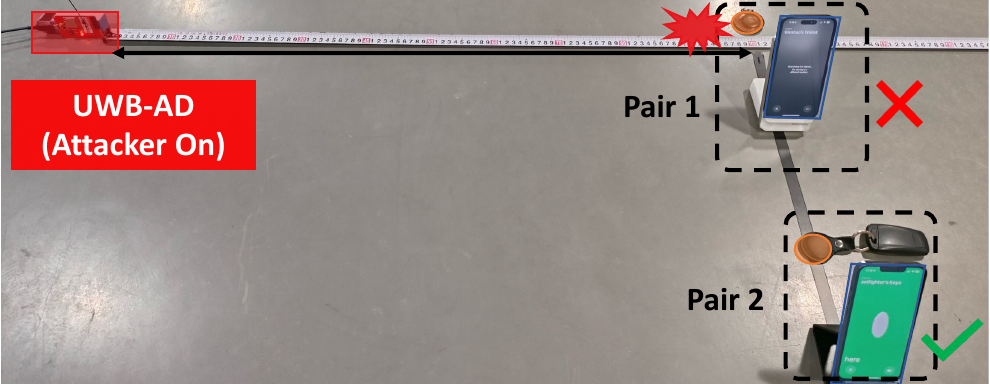}
        \caption{\zyz{Selective attack targeting a specific pair.}}
        \label{fig:airtag2}
    \end{subfigure}
     \vspace{1pt}
    \begin{subfigure}[b]{0.33\linewidth}
        \centering
        \includegraphics[width=1\textwidth]{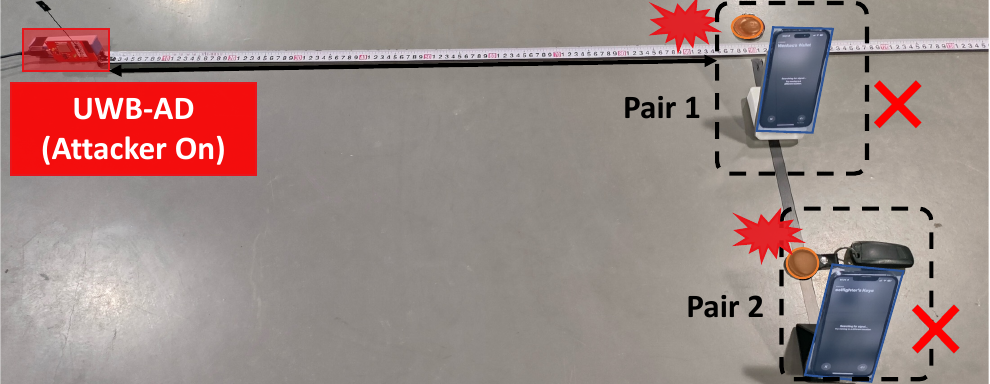}
        \caption{\zyz{Attack two pairs simultaneously.}}
        \label{fig:airtag3}
    \end{subfigure}
     \vspace{-22pt}
    \caption{\zyz{Attack commercial asset tracking devices (e.g., iPhone 14 and AiTag pairs). \projname{} can not only simultaneously disable the ranging capability of two paired iPhone-AirTag, but also selectively attack a targeted pair while leave the other UWB devices unaffected.}}\label{fig:airtag}
    \vspace{-10pt}
\end{figure*}

\subsection{Attack Case Studies}\label{sec:case_study}
We evaluate the performance of \projname{} in attacking real-world UWB applications with commercial UWB ranging systems, including asset tracking systems (iPhone and AirTag), indoor positioning systems (\zyz{three base stations and one localization tag}), and PKES (\zyz{newly released cars with UWB enabled}). Specifically, we use the maximum output power of \projname{} for jamming attacks. Before launching a real attack, \projname{} automatically sniffs the packet structure of the victim devices and decides the best attack delay.

\subsubsection{Attack Asset Tracking Systems}\label{sec:Asset Tracking}
\zyz{
AirTags are widely used for asset tracking and can be located by a paired iPhone through the ranging capabilities of the internal U1 UWB chips. As demonstrated in Figure~\ref{fig:airtag1}, two AirTags can be successfully located by their corresponding paired iPhones. These AirTags can be easily attached to items like wallets, keys, and bags, allowing users to effortlessly track their belongings. However, as shown in Figure~\ref{fig:airtag2}, under the attack of \projname{} (attack the final packet from the iPhone to the AirTag), the ranging session between the first iPhone-AirTag pair is fully disrupted, even when the iPhone is as close as $7cm$ to the AirTag, while \projname{} is positioned $100cm$ away from the AirTag. Interestingly, the ranging capability of the second iPhone-AirTag pair remains unaffected.
This is not surprising since \projname{} can only jam the ranging packets of the target iPhone-AirTag pair by carefully choosing a proper attack delay, while ignoring the ranging packets of other UWB devices.
This selective disruption makes \projname{} more imperceptible compared to full-band jamming, which would disrupt all UWB sessions simultaneously. Furthermore, as illustrated in Figure~\ref{fig:airtag3}, \projname{} can also disrupt the ranging sessions of two or more iPhone-AirTag pairs simultaneously, given that their attack delays can be precisely measured.
}

\begin{figure}[t]
    \centering
    \includegraphics[width=0.9\linewidth]{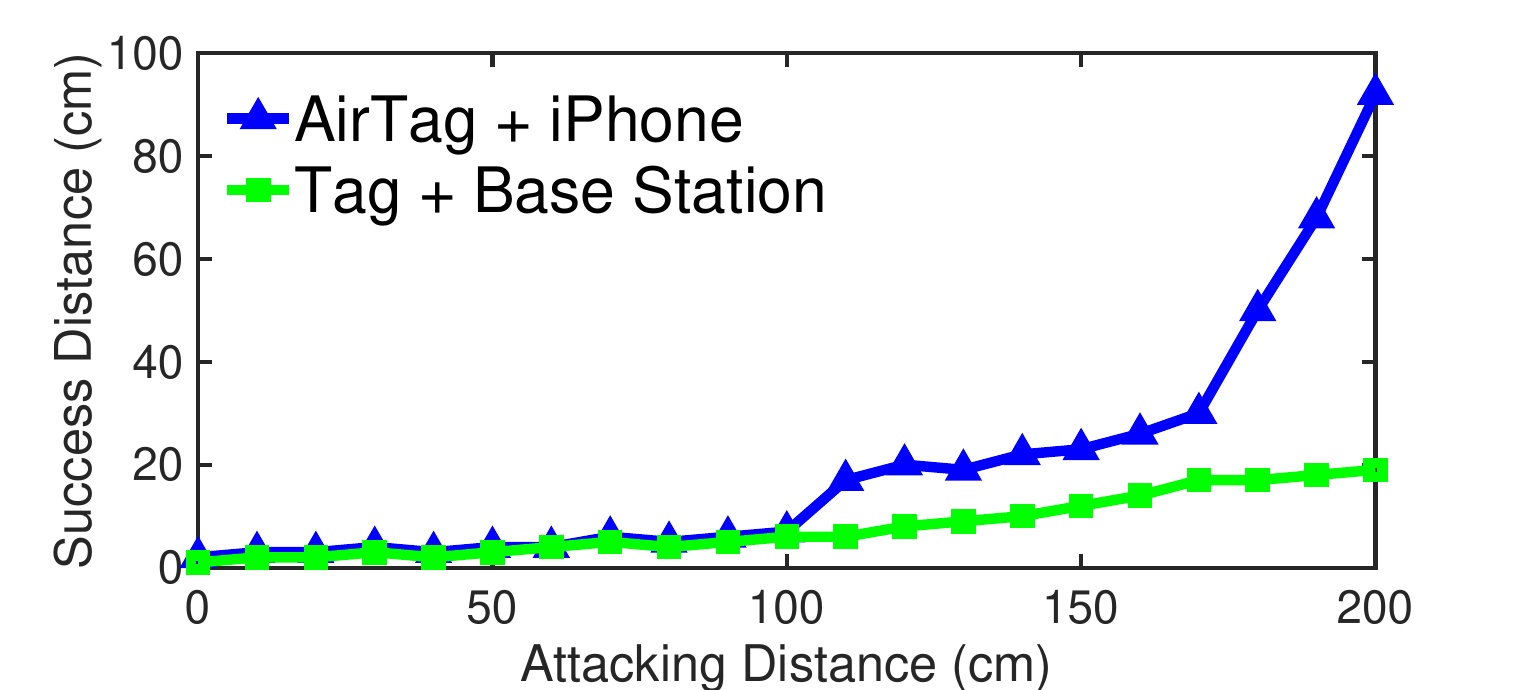}
    \vspace{-10pt}
    \caption{\zyz{Impact of the attacking distance (between the attacking device and the victim device) on the minimum success range (between an AirTag and iPhone, or a localization tag and the base station) at which \projname{} can fully disrupt the ranging sessions. Note that \projname{} operates effectively whenever the communication distance between paired victim UWB devices exceeds this minimum success range.}}
    \label{fig:iphoneandairtag}
    \vspace{-15pt}
\end{figure}

\zyz{
Furthermore, we aim to investigate the impact of the attacking distance—the distance over which the attack packets from \projname{} travel to the victim device—to assess the feasibility of \projname{} attacks against commercial products like the iPhone and AirTag. As direct tuning of the output power of UWB signals from iPhone is not possible, we adjust the distance between the iPhone and the AirTag and record the minimum distance at which \projname{} can fully disrupt the ranging sessions, denoted as "success range". Notably, the closer the paired UWB devices are, the stronger the signal of the legitimate ranging packets. Therefore, if the distance between the paired UWB devices is larger than the success range, \projname{} can completely disrupt their ranging sessions. The blue line in Figure~\ref{fig:iphoneandairtag} illustrates how the success range changes as the attacking distance increases from $0m$ to $2m$ for the iPhone and AirTag pair. As we can see, when the attacking distance is around $1m$ (i.e., \projname{} is placed $1m$ away from the AirTag), the iPhone should be within as close as $7cm$ of the AirTag to re-establish UWB ranging. This demonstrates the practical feasibility of \projname{} attacks on commercial products in real-world scenarios, which is benefiting from the power amplification module to boost the output power.
}

\subsubsection{Attack Indoor Localization Systems}
\zyz{
Next, we conduct a \projname{} attack against a commercial indoor positioning system~\cite{taobao} (equiped with Qorvo DW3000 series chips), which includes three base stations (responders) and a localization tag (initiator). Such systems are commonly used in warehouses, hospitals, and prisons to prevent unauthorized access. The core principle of this indoor localization system is triangulation localization, which calculates the distances between the tag and each base station and determines the true position by finding the intersection of three circles~\cite{tekdas2010sensor}. As a result, disrupting the ranging sessions between the tag and any base station will lead to a failure in localization.
}

\begin{figure}
    \begin{subfigure}{\linewidth}
        \centering
        \includegraphics[height=3.2cm]{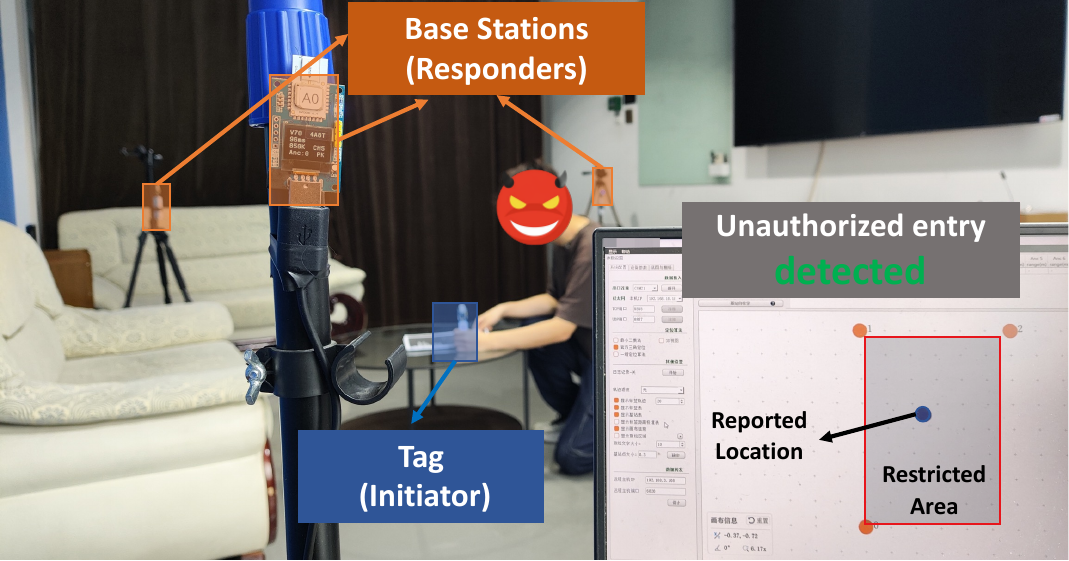}
        \caption{\zyz{Unauthorized entry triggers a detection alert.}}
        \label{fig:location1}
        \vspace{5pt}
    \end{subfigure}
    \begin{subfigure}{\linewidth}
        \centering
        \includegraphics[height=3.2cm]{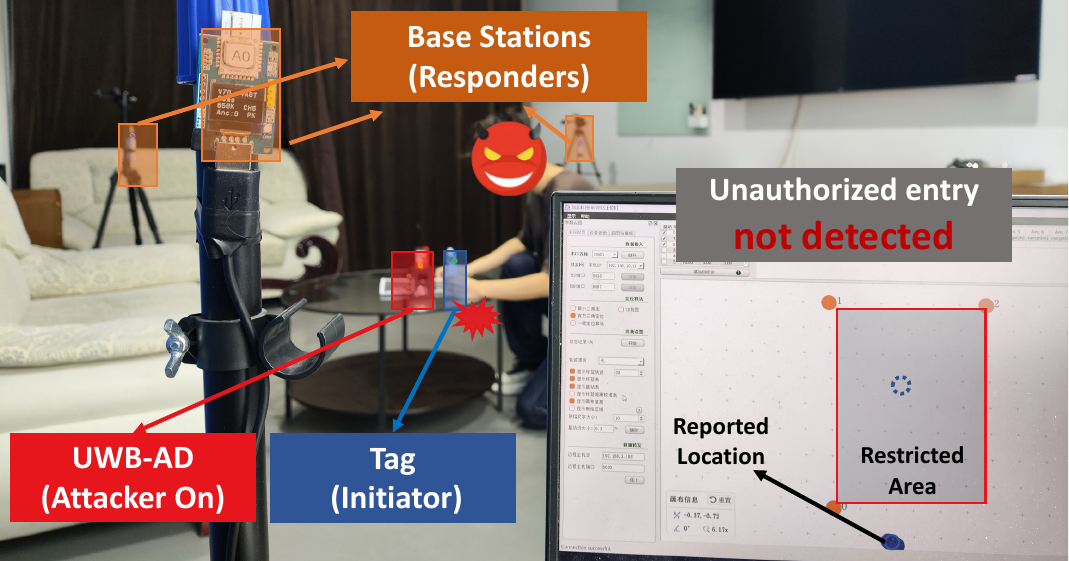}
        \caption{\zyz{Unauthorized entry undetected with \projname{} attack.}}
        \label{fig:location2}
    \end{subfigure}
    \vspace{-20pt}
    \caption{\zyz{Attack a commercial indoor positioning system, which uses $3$ base stations to localize the tag to prevent unauthorized entry to the restricted area.}}\label{fig:location}
    \vspace{-10pt}
\end{figure}

\zyz{
In our experiment, as illustrated in Figure~\ref{fig:location}, we simulate a scenario where an adversary, disguised as a legitimate worker, wears a non-removable localization tag. This setup ensures real-time monitoring of the physical locations of workers, and any unauthorized entry into a restricted area should trigger an alert, as shown in Figure~\ref{fig:location1}. However, if the adversary activates \projname{} to disrupt the ranging sessions between the tag and any base station before entering the restricted area (attack the response packet from one of the base stations to the tag), the system will fail to update the localization even when unauthorized entry occurs, as shown in Figure~\ref{fig:location2}. It's important to note that simply increasing the number of base stations (e.g., using $4$ or more base stations) does not effectively counteract \projname{}. The attack can still disrupt the ranging sessions between the localization tag and additional base stations (e.g., at least $2$ or more base stations), as it knows the exact time delays of each base station. 
}

\zyz{
We also evaluated how the success range changes with increasing attacking distance, as indicated by the green line in Figure~\ref{fig:iphoneandairtag}. When the attacking distance is below $1m$, the minimum success distance between the localization tag and the base station pair is similar to that of the iPhone and AirTag pair. However, when the attacking distance exceeds $1m$, the localization system becomes more susceptible to \projname{} attacks, as it requires a much smaller communication range to re-establish UWB ranging. This increased vulnerability is likely due to the lower signal power of the localization system, making it more susceptible to the injected attack packets.
}

\subsubsection{Attack Commercial Vehicles}\label{sec:case_pkes}

\zyz{
Passive Keyless Entry and Start (PKES) systems enbale users to unlock and start their vehicles by simply approaching the vehicle with key fobs or paired smartphones. Traditionally, these systems rely on narrow-band radio signals to detect if the key was within the vehicle's communication range. This method, however, is vulnerable to relay attacks~\cite{PKE23}, where adversaries could extend the communication range with long-range channels. To counter such issues, many high-end vehicles from manufacturers like BMW~\cite{BWMUWB1}, Tesla~\cite{TeslaUWB}, and NIO~\cite{NIOuwb1}, have upgraded their PKES systems to incorporate UWB technology, which rejects relay attacks by accurately verifying the physical distance between the car and the key fob (or the paired smartphones).
}

\zyz{
Nevertheless, \projname{} makes it possible again to attack these modern UWB-equipped PKES systems. To be specific, PKES systems separate the authentication message of the car key (via the narrow-band radio signals) from the physical distance bounding (via UWB). When a car receives an authentication message, it typically confirms the key's proximity by referring to the latest UWB ranging results.
However, \projname{} blocks the ranging sessions, causing the vehicle to rely on outdated distance data and wrongly assume the key is near or far. The \projname{} hardware can be discreetly installed in hidden locations on the vehicle, such as the undercarriage. The attacking device then captures the physical layer structure of UWB packets and determines the accurate timing to transmit jamming packets to disrupt the ranging sessions. Our experiments, conducted across various vehicle brands (which remain undisclosed due to ethical concerns), reveal two primary consequences of the \projname{} attack on PKES systems.
}

\begin{figure}
    \begin{subfigure}{\linewidth}
        \centering
        \includegraphics[height=3.2cm]{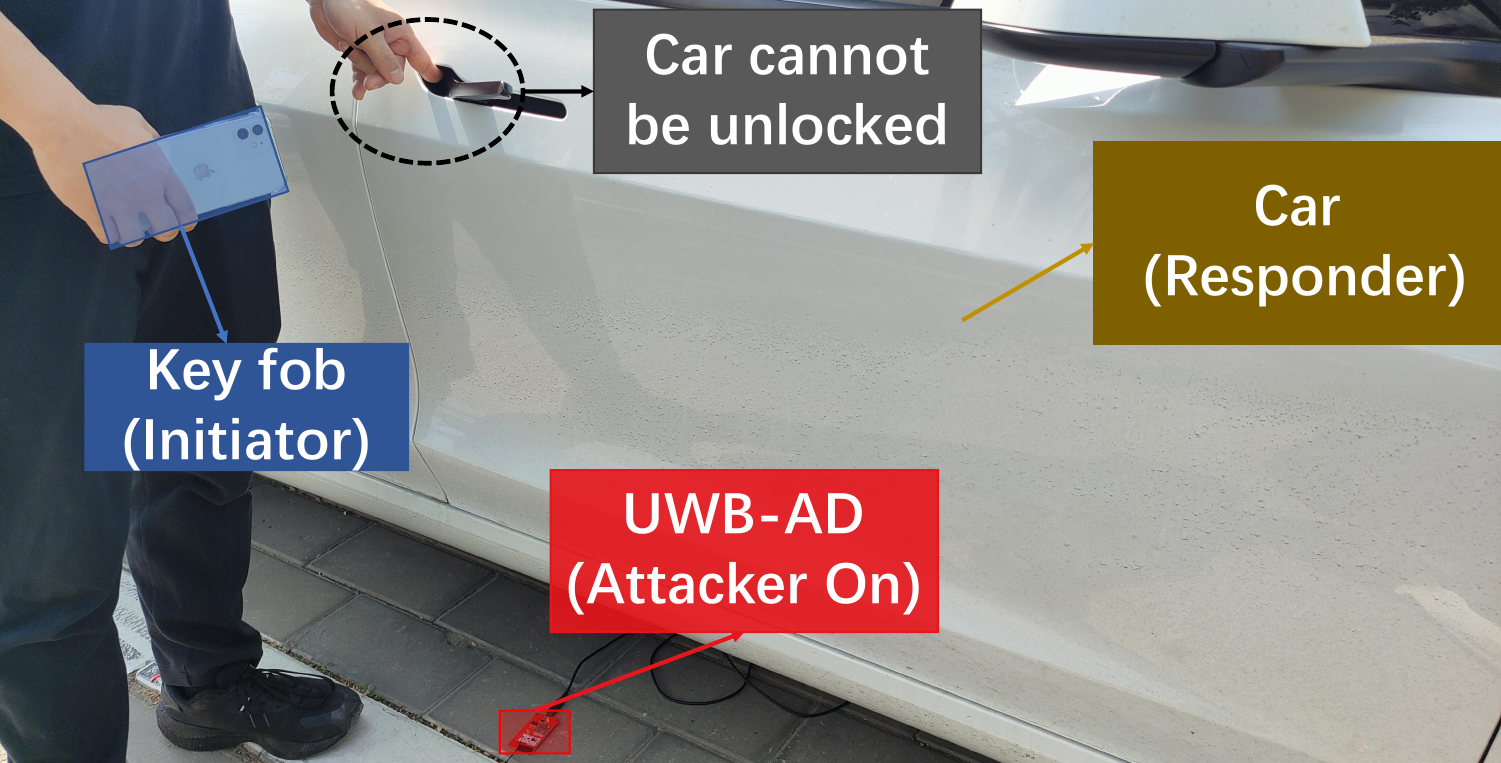}
        \caption{\zyz{Unable to open the car door under \projname{} attack.}}
        \label{fig:PKES1}
        \vspace{5pt}
    \end{subfigure}
    \begin{subfigure}{\linewidth}
        \centering
        \includegraphics[height=3.1cm]{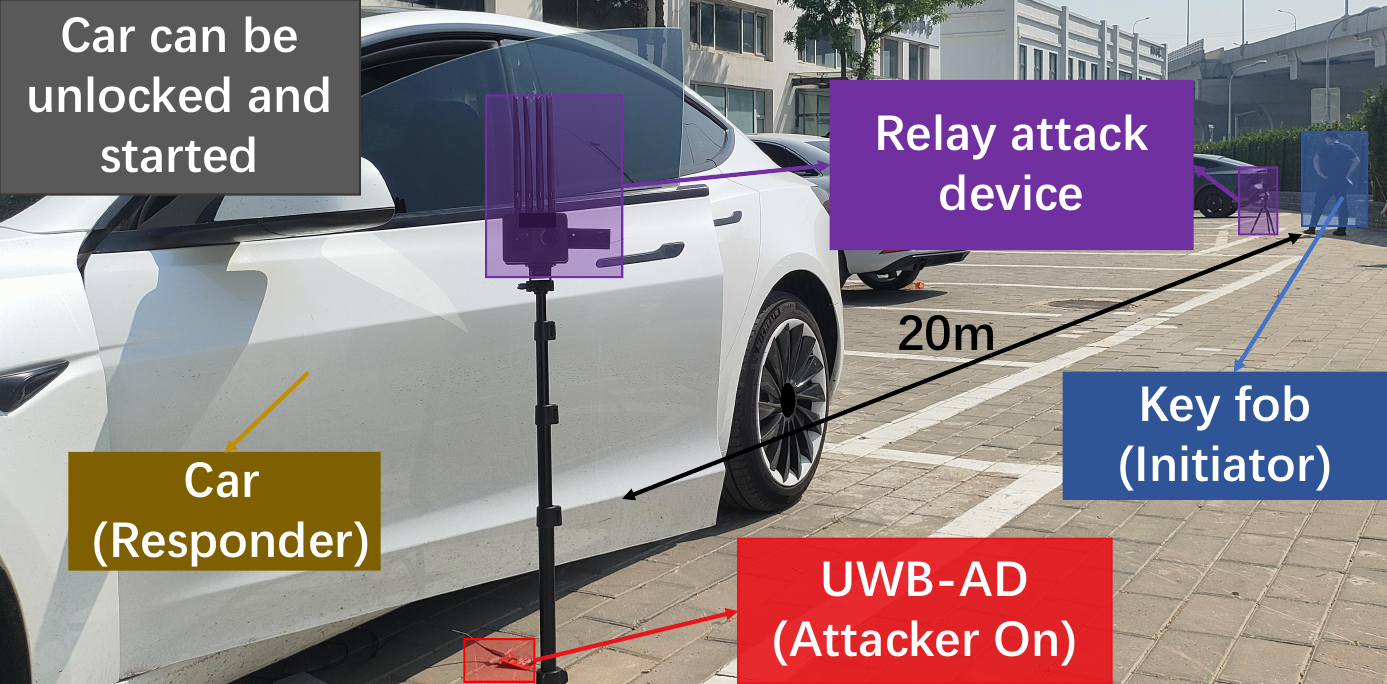}
        \caption{\zyz{Unauthorized vehicle access and theft after user leaves.}}
        \label{fig:PKES2}
    \end{subfigure}
    \vspace{-20pt}
    \caption{\zyz{Attack commercial UWB-equiped PKES systems. \projname{} makes it possible again to illegally unlock/lock and start the car, even when the PKES system is integrated with UWB adhering to 4z standard. }}
    \label{fig:15}
    \vspace{-23.1pt}
\end{figure}

\zyz{
\textbf{Unable to Open Car Door:} As demonstrated in Figure~\ref{fig:PKES1}, when a user approaches the vehicle from a distance (e.g., around $20m$) with a paired smartphone or key fob, the car doors are assumed to unlock automatically when the user is within around $5m$. Under the \projname{} attack, however, the door remains locked even when the user is within close proximity and attempts to pull the door handle to trigger the authentication message manually. This failure persists even after prolonged attempts to pull the door handle (e.g., around $5$ minutes), but the door unlocks immediately once \projname{} is deactivated, confirming the attack's ability to completely block UWB communications within modern PKES systems.
}

\zyz{
\textbf{Unauthorized Vehicle Access and Theft:} As shown in Figure~\ref{fig:PKES2}, when a user exits the vehicle and walks away, the doors should automatically lock once a certain distance is reached (e.g., $5m$). However, under the \projname{} attack, the doors remain unlocked, even if the user moves $20m$ away. This is not surprising because the car wrongly assumes the car key is still nearby, based on the outdated ranging data. Furthermore, if the user manually locks the car before leaving, adversaries could potentially relay the authentication message to unlock the car and start the engine. Such relay attacks can succeed within $2$ minutes since we activate \projname{}. This is possibly due to that the car retains old distance data for about $2$ minutes before requiring a new successful ranging session. This window provides ample opportunity for adversaries to steal the car. Because within $2$ minutes, the user can walk a considerable distance, remaining oblivious to the theft, thereby posing a significant security risk in practice.
}

\zyz{
The evolution of the PKES system follows the standards set by the Car Connectivity Consortium (CCC), with UWB-equipped systems first introduced in 2021~\cite{CCC}. Consequently, these security vulnerabilities could potentially impact a wide range of vehicles adhering to the latest CCC standards. Different car manufacturers, or even different models within the same brand, may have slight variations in the design of their PKES systems. These variations could include the proximity distance for detection, the duration for retaining outdated distance data, and the control logic, etc. Nevertheless, \projname{} has demonstrated its capability to successfully disrupt the UWB ranging in PKES systems across all tested vehicles (again, due to ethical concerns, the specific brands and models tested are not disclosed). Therefore, we recommend that car manufacturers consider the potential for UWB ranging failures and the inability to update distances in real-time when designing their PKES systems. This consideration is crucial for minimizing security threats as much as possible within the limitations of current UWB ranging protocols. We propose a potential countermeasure in Section~\ref{sec:counter} for current PKES systems.
}

\section{Discussion}
\label{sec:Discuss}


\subsection{Attack the upcoming IEEE 802.15.4ab} 

\subsubsection{Specific improvements to 802.15.4ab}
\zyz{
UWB chips have been adopted in many small devices such as smartphones, smartwatches, and tags, which lack high-gain antennas~\cite{ab1,Apple2,ab40}, thereby limiting the link budget.}
\zyz{
Therefore, the upcoming IEEE 802.15.4ab standard adopts the Narrowband Assisted Multi-Millisecond Ultra-Wideband (NBA-MMS-UWB) ranging proposal~\cite{MMSUWB} to improve the link budget. This proposal spreads multiple high-powered signal segments more sparsely at fixed intervals (e.g., 1ms or 2ms), thereby generating a longer effective packet. This allows the transmitter to increase the signal energy without violating the legal limit on average spectral power density, thus improving the overall link budget~\cite{4ab}.}

As demonstrated in Figure~\ref{fig:4ab}, Clock synchronization for NBA-MMS-UWB can utilize NB signals or continue to use the SYNC and SFD fields from 4z. For ranging functionality, Ranging Sequence Fragments (RSFs) and Ranging Integrity Fragments (RIFs) are used. RSFs are designed to establish ToF information and have a similar structure to the SYNC structure in the 4z standard, using CIR to receive the signal. On the other hand, RIFs are used for ToF verification to prevent distance-reducing attacks, which is similar to the function of STS in 4z. In addition, RIFs can also be used to establish and verify ToF simultaneously, and the specific design of the RIF is still under discussion at the time of writing. One of the proposed schemes is to continue using the STS waveform from 4z. This approach involves verifying ToF based on CIR, which can be obtained by cross-correlating the input signal with a local STS template.
\zyz{In summary, the current 4ab proposal maintains the SYNC structure and continues to use CIR for reception, with fixed packet intervals, which suggests that \projname{} could also be effectively applied to the upcoming 4ab standard.}

\subsubsection{Possibilities to attack 802.15.4ab}

\begin{figure}
    \centering
    \includegraphics[width=1\linewidth]{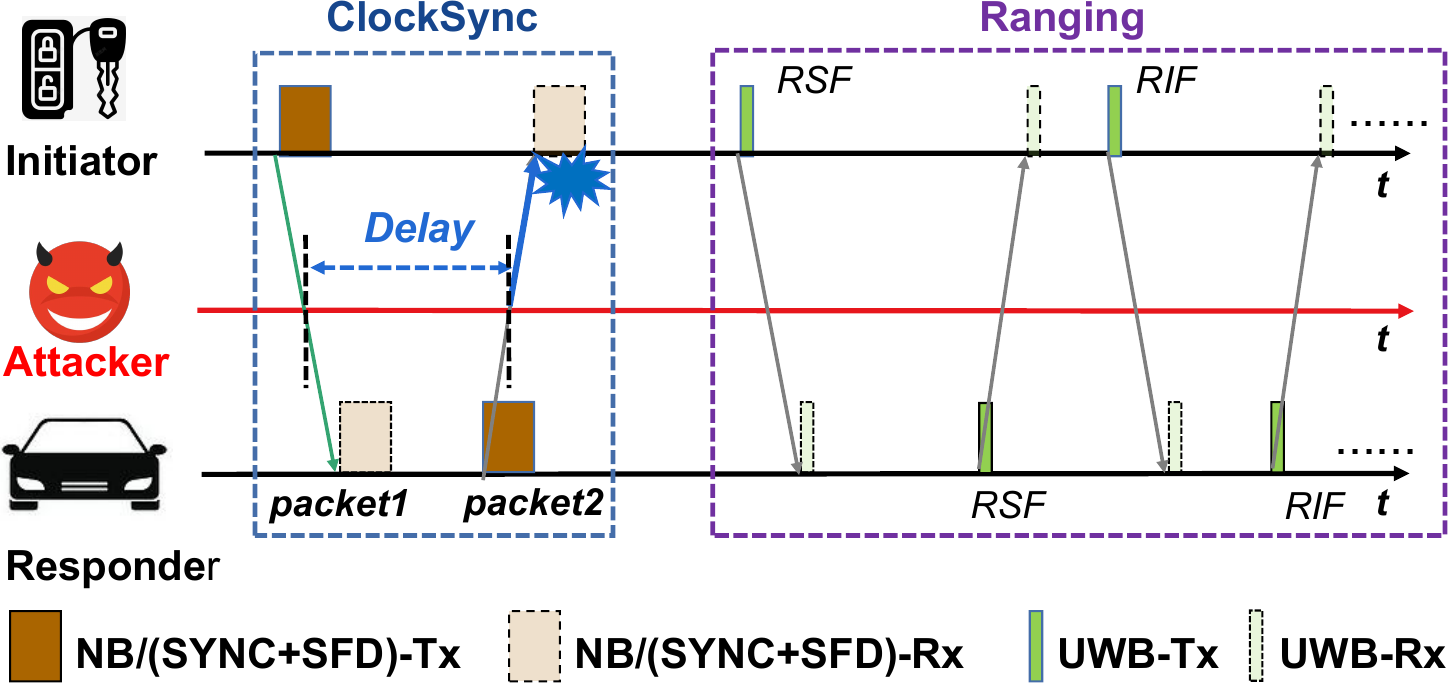}
    \vspace{-20pt}
    \caption{\zyz{Illustration of the 4ab proposal and the applicability of \projname{} in attacking UWB devices adhering to the upcoming 4ab standard.}}
    \label{fig:4ab}
    \vspace{-15pt}
\end{figure}

\zyz{Figure \ref{fig:4ab} illustrates the clock synchronization and ranging process between the initiator and the responder. Clock synchronization utilizes a bidirectional method, where the initiator sends $packet 1$ to the responder, who then sends $packet 2$ back. This approach helps mitigate asymmetric delays and frequency shift in unidirectional synchronization, thereby improving synchronization accuracy and system stability. The primary strategy of our \projname{} attack, as potentially applied to the 4ab proposal, involves measuring the timestamp of $packet 1$ and subsequently jamming $packet 2$ to disrupt the clock synchronization process. By using the timestamp of $packet 1$ as an anchor, we can then transmit attack packets after a fixed time delay when the initiator receives $packet 2$, inducing errors in clock synchronization and subsequently blocking the ranging sessions.}

\subsection{Countermeasures}\label{sec:counter}

\zyz{
To secure the UWB ranging in the upcoming IEEE 802.15.4ab standard, we first propose a countermeasure against \projname{} from the perspective of weakness of the UWB standard. Additionally, for vulnerabilities in the existing PKES systems, we also propose a practical countermeasure based on the design of PKES systems.}

\zyz{
We recommend introducing pseudo-random delays $T_{random}$ to the ranging packets, preventing \projname{} from predicting the arrival time of future packets. Given that the time interval between two ranging packets (e.g., $3 ms$) is significantly longer than the packet duration $T_0$ (e.g.,  $0.2 ms$), ensuring that the random $|T_{random}| \geq T_0$ results in no overlap between the attack packet and the legitimate packet. This arrangement ensures that regardless of the transmitting energy, the success rate of conducting \projname{} cannot approach nearly $100\%$, making it impractical to disrupt every ranging packet and ensuring that some packets can successfully update the true physical distance.
}

\zyz{
For commercial PKES systems utilizing UWB ranging, we recommend automatically denying access if no successful ranging occurs within a brief period, such as $30$ seconds. This contrasts with current PKE systems that retain old distance data for around $2$ minutes (refer to Section~\ref{sec:case_pkes}). Implementing this countermeasure involves simple modifications to the PKES system's control logic. However, such changes could negatively impact user experience, particularly if the UWB signal is blocked by obstacles, like when an owner is holding a metal object or standing behind a pillar. This might lead to the car automatically locking quickly, while other passengers might still need to access the vehicle. Consequently, there is a need to balance the accessibility of historical data and practical usage scenarios.
}

\subsection{Limitations and Future Work}
\zyz{
While \projname{} shows promise in disrupting commercial HRP UWB (Ultra-Wide Band) ranging systems, there are still some challenges to address. First, devices that use Low-Pulse-Rate (LRP) UWB chips may emit stronger pulse energy, which means our attack would need more power to jam effectively, potentially reducing the distance from which we can launch an attack. Second, if there are other UWB devices nearby with different packet structures, \projname{} might struggle to correctly identify the right signals during packet sniffing. As demonstrated in Sec.~\ref{sec:case_study}, \projname{} works well during both packet sniffing and attacking phases when all devices have the same packet structure, like the two pairs of iPhone-AirTag and the three localization base stations. But when devices use different packet structures, \projname{} might pick up the wrong signals because of interference. Despite this, once \projname{} successfully completes packet sniffing, it can launch attacks effectively even in the presence of interference. The packet sniffing phase could be potentially improved in two ways: (i) by identifying devices based on the strength of their signals, and (ii) by using \projname{}'s ability to be both effective and imperceptible to wait for a clear duration without interference before starting packet sniffing. We plan to keep the above challenges as our future work.
}

\section{Related work}
\label{sec:related}
Our work intersects with the following areas:

\textbf{Jamming Attacks: } 
\zyz{
The ultra-wide bandwidth of UWB renders it immune to narrow-band jamming~\cite{immunejamming}. In contrast, full-band jamming, which involves transmitting Gaussian white noise across the entire bandwidth, requires expensive, bulky, and customized hardware~\cite{keysight}. Constant emission of jamming packets and the uncontrolled range of indiscriminate attacks make full-band jamming easily detectable~\cite{fullbanddetect}.
To address these issues, \cite{dospaper} introduced two types of jamming attacks: (i) resource consumption interference, which aims to deplete system resources such as battery life and network bandwidth by transmitting numerous attack packets~\cite{5480628}. However, the continuous dispatch of attack packets is easily detected; (ii) preemptive transmission of a single legitimate signal to mislead the system into halting the ranging process~\cite{preemptivejamming}. This tactic is ineffective against the 4z protocol due to the pseudo-random characteristics of its STS field, which makes fabricating a deceptive signal impractical.
\projname{} is developed using COTS UWB chips and reacts by jamming attack packets when the victim devices receive legitimate packets, thereby disrupting the ranging sessions. This attack is effective, imperceptible, and low-cost. Moreover, it can leverage timing information to selectively target one or more UWB devices without impacting others, a strategy unachievable with full-band jamming. This capability significantly enhances the covert nature of the \projname{} attack and broadens the scope of potential attack case studies.
}

\textbf{Distance Reduction Attacks:} The Cicada attack, introduced in~\cite{poturalski2010cicada}, was the first distance reduction attack method.
In the 4a standard, the SYNC field serves for both clock synchronization and ToF calculation.~\cite{poturalski2010cicada} showed via simulation that injecting malicious signals into this field can reduce range.
 However, in the latest 4z standard, ToF calculation shifts from SYNC to the new STS field. Therefore, Singh et al.~\cite{singh2021security} designed distance reduction methods for attacking the STS field based on the Cicada attack on the SYNC field, namely Cicada++ and Adaptive. 
 Their simulations showed that these attacks could affect receiver ranging, with success rates between $7\%$ to $91\%$.
However, they did not test these attacks on actual UWB chips or confirm their reliability on existing hardware.
To explore how these methods can attack real products,~\cite{leu2022ghost} introduced a similar but more practical attack called Ghost Peak and conducted real distance reduction attacks on the Apple U1 chip. 
It had a low $4\%$ success rate in practical attacks due to its reliance on injecting random noise into the STS field, which is not reliable as filters can easily remove the distorted data~\cite{welch1995introduction, justusson2006median}.
Furthermore, this attack can be thwarted by comparing timestamps measured from multiple fields, as the random ToF reduction in the STS field is challenging to reproduce~\cite{luo2023secure,firaAccurateRanging,joo2023protecting}.
To increase the success rate of attacks, recent research~\cite{anliker2023time} has proposed clock disruption-based MD and SaA attacks. Notably, the MD attack consistently reduces the distance from $10 m$ to $0 m$ on development kits, yet its efficacy is unreliable in practical scenarios. This limitation stems from its effectiveness being confined solely to the SS-TWR ranging method, which is notoriously susceptible to clock drift. SaA's effectiveness, aimed at the unreleased ab standard, remains untested in practice due to the absence of 4ab-based products.
The main idea behind our \projname{} attack is not to reduce the distance but to use jamming attacks to prevent distance information from updating, causing the attacked system to continuously use outdated distance data, achieving a similar effect to distance reduction attacks. Through experiments, \projname{} has proven to be highly reliable in real-world scenarios and can consistently succeed in attacking products based on the 4z standard.

\section{Conclusion}
\label{sec:conclusion}

In this paper, we explore the potential for disrupting the ranging sessions of commercial UWB ranging systems adhering to the widely adopted IEEE 802.15.4z standard, with potential applicability to the forthcoming IEEE 802.15.4ab standard. We discover the normalized cross-correlation process at the receiver side could be leveraged as a vulnerability for jamming attacks, rendering effective, imperceptible, and low-cost field-level jamming. Subsequently, we develop \projname{} utilizing readily available COTS UWB chips, resulting in a less imperceptible and reactive system. \projname{} is capable of preparing attacks for unseen UWB products without manual intervention. Our experiments demonstrate the tangible and effective impact of \projname{} on commercial UWB ranging systems from the three leading UWB chip vendors,  including Apple, NXP, and Qorvo. These findings have prompted internal security incident response procedures at several companies.

\begin{acks}
 We sincerely thank the anonymous reviewers for their insightful and helpful comments. This work was supported in part by National Key R\&D Program of China (2022YFB4501200), National Natural Science Foundation of China (62332004, U2336204), Sichuan Natural Science Foundation (2023NSFSC1963, 24ZNSFSC0038), Key Research Funds of Sichuan Province (24GJHZ0225), and Fundamental Research Funds for Chinese Central Universities (ZYGX2021J018).
\end{acks}

\bibliographystyle{ACM-Reference-Format}
\normalem
\bibliography{sample-base}


\end{document}